\documentclass[12pt]{article}
\usepackage{amsmath,amssymb,booktabs}

\usepackage{cancel}

\makeatletter

\@addtoreset{equation}{section}
\makeatother

\usepackage[pdftex]{graphicx,color}
\usepackage{epstopdf}
\usepackage[bookmarks=true,bookmarksnumbered=true,bookmarkstype=toc]{hyperref} 

\usepackage{color}

\definecolor{green}{cmyk}{1,0,1,0}

\usepackage{ulem}  % \sout{ text }
\begin{document}

\title{\vbox{
\baselineskip 14pt
\hfill \hbox{\normalsize KUNS-2395}\\
\hfill \hbox{\normalsize KYSHU-HET-133}\\
\hfill \hbox{\normalsize YITP-12-19}} \vskip 1.7cm
\bf TeV scale mirage mediation \\%
in NMSSM\vskip 0.5cm
}
\author{%
Tatsuo~Kobayashi$^{1}$,\ 
Hiroki~Makino$^{2}$, \
Ken-ichi~Okumura$^{2}$, \\
Takashi~Shimomura$^{3}$, \ and  
Tsubasa~Takahashi$^{4}$
\\*[20pt]
$^1${\it \normalsize
Department of Physics, Kyoto University, Kyoto 606-8502, Japan} \\
$^2${\it \normalsize 
Department of Physics, Kyushu University, Fukuoka 812-8581,
Japan} \\
$^3${\it \normalsize Department of Physics, Niigata University,~Niigata 950-2181, Japan} \\
$^4${\it \normalsize Yukawa Institute for Theoretical Physics, Kyoto University,
Kyoto 606-8502, Japan}
}

\date{}

\maketitle
\thispagestyle{empty}

\begin{abstract}
We study the next-to-minimal supersymmetric standard model.
We consider soft supersymmetry breaking parameters, 
which are induced by the mirage mediation mechanism 
of supersymmetry breaking.
We concentrate on the mirage mediation, where 
the so-called mirage scale is the TeV scale.
In this scenario, we can realize 
%%KO%% the effective $\mu$-term 
the up-type Higgs soft mass
%%KO%%
of ${\cal O}(200)$ GeV, while other masses such as 
gaugino masses and stop masses are heavy such as 1 TeV or more.
%KO
Cancellation between the effective $\mu$-term and
the down-type Higgs soft mass ameliorates
the fine-tuning in the electroweak symmetry breaking 
even for $\mu={\cal O}(500)$ GeV. 
%KO
%The lightest CP-even Higgs mass can be $115-130$ GeV.
%The lightest CP-even Higgs mass can reach 125 GeV with $\tan\beta\simeq 3$.
%%KO>
The mixing between the doublet and singlet Higgs bosons is suppressed by $(\lambda/\kappa)\tan^{-1}\beta$.
Then the lightest doublet Higgs mass naturally reaches 125 GeV lifted by the new quartic coupling.
%%<KO
The higgsino and singlino are light and their linear 
combination is the lightest superparticle.
\end{abstract}

\newpage

\setcounter{footnote}{0}

\section{Introduction}

Supesymmetric extension is a good candidate for physics 
beyond the standard model (SM).
The minimal supersymmetric standard model (MSSM) 
is the simplest extension.
The MSSM is quite interesting because of its minimality 
and its detailed studies have been done for several 
aspects.

However,  the MSSM has the fine-tuning problem.
Within the framework of the MSSM, 
the $Z$-boson mass, $m_Z$, is obtained as 
\begin{equation}\label{eq:Mz-Hu}
\frac{m_Z^2}{2} \approx -m^2_{H_u}-|\mu|^2,
\end{equation} 
where $m^2_{H_u}$ is the soft supersymmetry (SUSY) breaking 
scalar mass squared of 
the up-sector Higgs field and $\mu$ is the supersymmetric mass.
The radiative corrections on $m^2_{H_u}$ are obtained as 
$m^2_{H_u} \sim -m^2_{\tilde t} \sim - M_3^2$, 
where $m_{\tilde t}$ and $M_3$ denote the stop and 
gluino masses, respectively.
In most of cases, the stop and the gluino masses are 
much larger than $m_Z$.
Thus, we need fine-tuning between $m^2_{H_u}$ and 
$|\mu|^2$ to realize the correct value of $m_Z$.
Furthermore, it is required that $m_{\tilde t} ={\cal O}(1)$ TeV 
%%KO> 
or larger
%%<KO
%in order to obtain the Higgs mass such as $m_h> 115$ GeV.
in order to obtain the Higgs mass such as $m_h\approx 125$ GeV
which is recently reported by ATLAS and CMS collaborations\cite{:2012gk,:2012gu}.
%%KO%%LHC result

The mirage mediation is one of 
the interesting mediation mechanisms of 
SUSY breaking \cite{Choi:2004sx,Choi:2005uz,Endo:2005uy}.
The mirage mediation is a mixture of 
the modulus mediation \cite{Kaplunovsky:1993rd} 
and the anomaly mediation \cite{Randall:1998uk}
with a certain ratio.
In particular, it was pointed out that 
the TeV-scale mirage mediation can 
ameliorate the above fine-tuning problem of 
the MSSM \cite{Choi:2005hd,Kitano:2005wc,Choi:2006xb}.
In the TeV-scale mirage mediation, 
the above radiative corrections on $m^2_{H_u}$ 
and the anomaly mediation contributions are canceled each other.
Then, the value of $|m^2_{H_u}|$ at the 
electroweak scale can be smaller than 
stop and gluino masses.
The TeV scale mirage mediation also leads several 
phenomenologically interesting aspects \cite{Falkowski:2005ck} because 
its SUSY particle spectrum is quite compressed.
%%KO%% B-\mu problem

The next-to-minimal supersymmetric standard model 
(NMSSM) is the extension of the MSSM by 
adding a singlet $S$ \cite{Fayet:1974pd} 
(see for review e.g. \cite{Ellwanger:2009dp}).
Here, we also impose the $Z_3$ symmetry.
The NMSSM does not have the $\mu$-term, 
$\mu H_u H_d$, in the superpotential, 
where $H_u$ and $H_d$ denote the up and 
the down-sector Higgs superfields, respectively.
On the other hand, 
the term $\lambda S H_u H_d$ 
is allowed in the NMSSM superpotential.
After $S$ develops its vacuum expectation value (VEV), 
the effective $\mu$-term is generated.
That gives us a solution for the so-called $\mu$-problem \cite{Kim:1983dt}.
%%KO
The NMSSM is also interesting in the light of
the recent indication of the relatively heavy Higgs boson
reported by ATLAS and CMS, % \cite{ATLAS:2012ae,Chatrchyan:2012tx},
endowed with an additional Higgs self-coupling. 
In the Higgs sector, the doublet Higgs and singlet fields 
mix each other.
Then, the Higgs sector in the NMSSM has a 
quite rich structure.

The NMSSM also leads the same relation as (\ref{eq:Mz-Hu}),
and the NMSSM has the fine-tuning problem similar to 
the one in the MSSM.
Thus, it is interesting to apply the TeV-scale 
mirage mediation to the NMSSM.
In this paper, we study the NMSSM with 
the soft SUSY breaking terms induced through 
the TeV-scale mirage mediation \cite{takahashi}.

This paper is organized as follows.
In section 2, we give a brief review on 
the mirage mediation, and the TeV scale mirage mediation.
In section 3, we apply the TeV scale 
mirage mediation to the NMSSM, and 
study its spectrum.
Section 4 is devoted to conclusion and discussion.
In Appendix A, we show explicitly 
initial conditions of soft parameters, 
which are induced through the mirage mediation  
in the NMSSM.
%In Appendix B, we show our method to evaluate 
%the Higgs boson masses in the NMSSM. 

\section{TeV-scale mirage mediation}

Here, we review briefly the mirage mediation \cite{Choi:2004sx}.
The mirage mediation is the mixture between 
the modulus mediation and the anomaly mediation.
Then, the gaugino masses are obtained as 
\begin{eqnarray}\label{eq:gaugino-mass}
M_a = M_0 +\frac{m_{3/2}}{8 \pi^2}b_a g_a^2,
\end{eqnarray}
where $g_a$ and $b_a$ are the gauge couplings 
and their $\beta$ function coefficients, and $m_{3/2}$ denotes 
the gravitino mass.
The first and the second terms in the right-hand side of 
Eq.~(\ref{eq:gaugino-mass}) correspond to 
the contributions due to the modulus 
mediation and the anomaly mediation, respectively.

Similarly, we obtain the so-called $A$-terms corresponding to the
Yukawa couplings, $y_{ijk}$, and the soft scalar masses $m_i$ as 
\begin{eqnarray}\label{eq:A-m}
A_{ijk}(M_{GUT}) &=& a_{ijk}M_0 - (\gamma_i + \gamma_j + \gamma_k)\frac{m_{3/2}}{8\pi^2}, \nonumber \\
m_i^2(M_{GUT}) &=& c_i M_0^2 - \dot{\gamma_i}(\frac{m_{3/2}}{8\pi^2})^2
				- \frac{m_{3/2}}{8\pi^2} M_0 \theta_i,
\end{eqnarray}
where
\begin{eqnarray}
%b_a &=& -3\mathrm{tr}(T_a^2(\mathrm{Adj})) + \sum_i \mathrm{tr}(T_a^2(\phi^i)), \nonumber \\
\gamma_i &=& 2\sum_a g_a^2 C_2^a(\phi^i) - \frac{1}{2} \sum_{jk} |y_{ijk}|^2, \nonumber \\
\theta_i &=& 4\sum_a g_a^2 C_2^a(\phi^i) - \sum_{jk} a_{ijk} |y_{ijk}|^2, \nonumber \\
\dot{\gamma_i} &=& 8\pi^2 \frac{d\gamma_i}{d \ln \mu_R}.
\end{eqnarray}
Here, $C_2^a(\phi^i)$ denotes the quadratic Casimir 
corresponding to the representation of the matter field $\phi^i$.
In addition, $a_{ijk}$ and $c_i$ parametrize the A-term and the scalar mass squared 
generated through the modulus mediation in the unit of the universal gaugino mass, $M_0$.
These coefficients are determined by modulus-dependence 
of the K\"ahler metric as well as the Yukawa coupling.
One can write $c_i$ as 
\begin{equation}
c_i = c_i^{(\rm tree)} + \delta c_i^{(\rm loop)}.
\end{equation}
Here, $c_i^{(\rm tree)}$ is calculated from the tree-level 
K\"ahler metric of the matter field $\phi^i$ and they 
are ratios of small integers including $0$ and $1$ 
\cite{Kaplunovsky:1993rd,Abe:2005rx,Choi:2006xb}.
In addition, $\delta c_i^{(\rm loop)}$ is obtained with 
the one-loop K\"ahler metric of the matter field, 
but such a loop correction to the K\"ahler metric 
depends on the detail of the ultraviolet-model and is hard to calculate (see e.g. Ref.~\cite{Choi:2008hn}).
That is, $c_i$ is ambiguous at the one-loop level, 
although such ambiguity is subdominant and less important 
in most of cases.
Here, we consider the case with 
\begin{equation}
a_{ijk} = c_i + c_j +c_k.
\end{equation}
We input the values of $c_i$ at $M_{GUT} =2\times 10^{16}$ GeV.

It is convenient to use the following parameter \cite{Choi:2005uz},
\begin{equation}
\alpha \equiv \frac{m_{3/2}}{M_0 \ln(M_{pl}/m_{3/2})},
\end{equation}
to represent the ratio of the anomaly mediation 
to the modulus mediation. Here $M_{pl}$ is the reduced Planck scale.

One of the interesting aspects in the mirage mediation 
is that the  above spectrum \eqref{eq:gaugino-mass} and 
\eqref{eq:A-m} has a special energy 
scale, that is, the mirage scale, 
\begin{equation}
M_{\rm mir} = \frac{M_{GUT}}{(M_{pl}/m_{3/2})^{\alpha/2}} .
\end{equation}
At this scale, the gaugino masses are obtained as \cite{Choi:2005uz},
\begin{equation}\label{eq:gaugino-mir}
M_a(M_{\rm mir}) = M_0.
\end{equation}
That is, the anomaly mediation contribution and the radiative 
corrections cancel each other, and 
the pure modulus mediation appears 
at the mirage scale.
Furthermore, the $A$-terms and the scalar masses squared
also satisfy\footnote{The scaler masses at the Mirage scale can be modified due to 
the U(1) tadpole contribution in the renormalization group running when the different values of 
$c_{H_u}$ and $c_{H_d}$ are chosen. However, such a modification is small and can be included in ambiguities of 
$c_i$ if couplings are small. See \cite{Choi:2005uz} for detailed discussions. We include this contribution in our numerical analysis.} 
\begin{equation}\label{eq:A-m-mir}
A_{ijk}(M_{\rm mir}) = (c_i+c_j+c_k)M_0, \qquad m^2_i(M_{\rm mir}) =c_iM_0^2,
\end{equation}
if the corresponding Yukawa couplings are small enough or 
if the following conditions are satisfied, 
\begin{equation}\label{eq:mir-condition}
a_{ijk}=c_i+c_j+c_k=1,
\end{equation}
for non-vanishing Yukawa couplings, $y_{ijk}$ \cite{Choi:2005uz}.

When $\alpha =2$, the mirage scale $M_{\rm mir}$ is 
around $1$ TeV.
Then, the above spectrum (\ref{eq:gaugino-mir}) and 
(\ref{eq:A-m-mir}) is obtained at the TeV scale.
That is the TeV scale mirage mediation scenario.
In particular, there would appear a large gap 
between $M_0$ and the scalar mass $m_i$ with $c_i \approx 0$.
We will apply the TeV scale mirage scenario to the NMSSM in the next section.

In the TeV scale mirage scenario, the stop mass squared 
becomes negative at high energy \cite{Lebedev:2005ge}, 
while it is positive 
at low energy below $10^{6}$ GeV.
Thus, the vacuum which breaks the electroweak symmetry
at the electroweak scale might be a local minimum, but instead  
there would be a color and/or charge breaking vacuum 
with  field values larger than $10^{6}$ GeV.
Here, we assume the thermal history of the Universe 
such that field values remain around the origin 
until the temperature reaches the electroweak scale.
In addition, we need to confirm that the tunnelling rate is 
small enough, i.e. less than the Hubble expansion rate.
In Refs.~\cite{Riotto:1995am}, it has been shown that 
such a rate is small enough, as long as 
the squark/slepton masses squared are vanishing or positive 
around $10^4$ GeV.
This condition is satisfied in our TeV scale mirage mediation
scenario.

\section{TeV scale mirage in NMSSM}

In this section, we apply the TeV scale mirage mediation scenario 
to the NMSSM.

\subsection{NMSSM}

Here, we briefly review on the NMSSM, in particular its Higgs sector 
before we apply the TeV scale mirage mediation scenario to the NMSSM.
In the NMSSM, we extend the MSSM by adding a singlet chiral multiplet $S$ 
and imposing a $Z_3$ symmetry.
Then, the superpotential of the Higgs sector is written as 
\begin{equation}
W_{\rm Higgs} = - \lambda S H_u H_d + \frac{\kappa}{3}S^3.
\end{equation}
Here and hereafter, for $S$, $H_u$ and $H_d$ we use the convention that 
the superfield and its lowest component are 
denoted by the same letter. 
The full superpotential also includes the Yukawa coupling terms 
between the matter fields and the Higgs fields, which are 
the same as those in the MSSM.

The following soft SUSY breaking terms are induced 
in the Higgs sector,
\begin{equation}
V_{\rm soft}= m^2_{H_u}|H_u|^2 + m^2_{H_d}|H_d|^2 
+m^2_S |S|^2-\lambda A_\lambda SH_uH_d + \frac{\kappa}{3}A_\kappa S^3+ h.c.
\end{equation}
Then, the scalar potential of the neutral Higgs fields is 
given as 
\begin{eqnarray}
V &=& \lambda^2|S|^2(|H^0_d|^2+|H^0_u|^2)+|\kappa S^2- \lambda
H^0_dH^0_u|^2 +V_D \nonumber \\
& & +m^2_{H_u}|H_u|^2 + m^2_{H_d}|H_d|^2 
+m^2_S |S|^2-\lambda A_\lambda SH_uH_d + \frac{\kappa}{3}A_\kappa S^3+ h.c.,
\end{eqnarray}
with
\begin{equation}
V_D = \frac18 (g^2_1 + g^2_2)(|H^0_d|^2 - |H^0_u|^2)^2,
\end{equation}
where $g_1$ and $g_2$ denote the gauge couplings of U(1)$_{\rm Y}$  
and SU(2).

The minimum of the potential is obtained by analyzing 
the stationary conditions of the Higgs potential,
\begin{subequations}
\begin{align}
\frac{\partial V}{\partial H^0_d} &= \lambda^2 v \cos\beta ( s^2 + v^2 \sin^2 \beta )
  - \lambda \kappa v s^2 \sin\beta + \frac{1}{4} g^2 v^3 \cos\beta \cos2 \beta \nonumber \\
 &\quad + m_{H_d}^2 v \cos\beta - \lambda A_\lambda v s \sin\beta = 0, \label{eq:1sub1} \\
\frac{\partial V}{\partial H^0_u} &= \lambda^2 v \sin\beta ( s^2 + v^2 \cos^2 \beta )
  - \lambda \kappa v s^2 \cos\beta - \frac{1}{4} g^2 v^3 \sin\beta \cos2 \beta \nonumber \\
 &\quad + m_{H_u}^2 v \sin\beta - \lambda A_\lambda v s \cos\beta = 0, \label{eq:1sub2} \\
\frac{\partial V}{\partial S} &= \lambda^2 s v^2 + 2 \kappa^2 s^3 - \lambda \kappa v^2 s \sin 2\beta
 + m_S^2 s - \frac{1}{2} \lambda A_\lambda v^2 \sin 2\beta + \kappa A_\kappa s^2 = 0, \label{eq:1sub3}
\end{align}
\label{eq:1}
\end{subequations}
where $g^2=g^2_1+g^2_2$.
Here, we denote VEVs as 
\begin{equation}
v^2 = \langle |H^0_d|^2 \rangle + \langle |H^0_u|^2 \rangle, \qquad \tan \beta = \frac{\langle H^0_u \rangle }{\langle H^0_d \rangle}, 
\qquad s = \langle S \rangle.
\end{equation}
Using the above stationary conditions, we obtain the $Z$ boson mass $m_Z^2=\frac12 g^2v^2$ as
\begin{equation}
m_Z^2 = \frac{1 - \cos 2\beta}{\cos 2\beta} m_{H_u}^2 - \frac{1 + \cos 2\beta}{\cos 2\beta} m_{H_d}^2
  - 2\mu^2, 
\end{equation}
%%%KO
%where $\tan\beta$ also satisfies,
%\begin{equation}
%\frac{\tan^2\beta+1}{\tan\beta}=\frac{m_{H_d}^2+m_{H_u}^2+2\mu^2+\frac{2\lambda^2}{g^2}m_Z^2}{\left(A_\lambda+\frac{\kappa}{\lambda}\mu\right)\mu}.
%\end{equation}
where $\mu = \lambda s$.
For $\tan\beta \gg 1$, this equation becomes
\begin{equation}
m^2_Z  \simeq - 2 m_{H_u}^2 + \frac{2}{\tan^2 \beta} m_{H_d}^2
  - 2\mu^2 \   .
\label{eq:2}
\end{equation}
This relation is the same as the one in the MSSM.
Indeed, when we neglect the second term in the right-hand side, 
the above relation is nothing but Eq.~(\ref{eq:Mz-Hu}).
Thus, the natural values of $|m_{H_u}|$ and $|\mu|$ would be 
of ${\cal O}(100)$ GeV.
Furthermore, the natural value of $|m_{H_d}|/\tan \beta$ would be 
of  ${\cal O}(100)$ GeV or smaller.
Alternatively, $|\mu|$ and $|m_{H_d}|/\tan \beta$ could be larger 
than ${\cal O}(100)$ GeV when 
$\mu^2$ and $m_{H_d}^2/\tan^2 \beta$ are canceled each other 
in the above relation at a certain level.
Even in such a case, $|m_{H_u}|$ would be naturally of ${\cal O}(100)$ GeV.
On the other hand, other sfermion masses as well as gaugino 
masses must be heavy as the recent LHC results suggested.
To realize such a spectrum, we apply the 
TeV scale mirage mediation in the next section, 
where we take $c_{H_u}=0$ to realize a suppressed value of 
$|m_{H_u}|$ compared with $M_0$.

\subsection{TeV scale mirage mediation in NMSSM}

Here, we study the TeV scale mirage mediation scenario in the NMSSM.
Soft SUSY breaking terms are obtained 
through the generic formulas (\ref{eq:gaugino-mass}) 
and (\ref{eq:A-m}) with taking $\alpha =2$.
For concreteness, we give explicit results of 
all the soft SUSY breaking terms  for 
the NMSSM in Appendix \ref{app:soft}.
We concentrate on the Higgs sector as well as gauginos and stops.

We consider the following values of $c_i$,
\begin{equation}\label{eq:ci}
c_{H_d}^{(\rm tree)}=1, \qquad  c_{H_u}^{(\rm tree)}=0, \qquad c_S^{(\rm tree)} = 0, \qquad c_{t_L}^{(\rm tree)}=c_{t_R}^{(\rm tree)}=\frac12, 
\end{equation}
for $H_d$, $H_u$, $S$, and left and right-handed (s)top fields, respectively.
This is the same assignment as the pattern II in Ref. \cite{Choi:2006xb} 
for the MSSM except for $c_S$.
Then, the soft parameters due to only modulus mediation 
contribution are given by 
\begin{eqnarray}
& & (A_t)_{\rm modulus}=(A_\lambda )_{\rm modulus}=M_0, \qquad (A_\kappa )_{\rm modulus} =0, \nonumber \\
& & (m^2_{H_d})_{\rm modulus}=M^2_0, \qquad (m^2_{\tilde t_L})_{\rm modulus} =
(m^2_{\tilde t_R})_{\rm modulus} =\frac12 M^2_0,  \\ 
& &  (m^2_{H_u})_{\rm modulus }=(m^2_S)_{\rm modulus}=0,   \nonumber 
\end{eqnarray}
when we neglect $\delta c_i^{(\rm loop)}$.
The above assignment of $c_i$ (\ref{eq:ci})
satisfies the condition, (\ref{eq:mir-condition}) for 
the top Yukawa coupling and the coupling $\lambda$, but 
not for the coupling $\kappa$.
However, we do not consider a large value of $\kappa$ 
to avoid the blow-up of $\kappa$ and $\lambda$ as will be shown later.
Thus, we obtain the following values,
\begin{eqnarray}\label{eq:soft-at-mirage-1}
& & A_t \approx A_\lambda \approx M_0, \\ \nonumber 
&  &  m^2_{H_d} \approx M^2_0, \qquad m^2_{\tilde t_L} \approx m^2_{\tilde t_R} \approx \frac12 M^2_0, 
\end{eqnarray}
up to ${\cal O}(\kappa^2/8\pi^2)$ at the TeV scale.
Note that $\delta c_i^{(\rm loop)}$ has negligible effects for these values.

Similarly, we can obtain the values of $A_\kappa$, $|m_{H_u}|$ and 
$|m_s|$ at the TeV scale, however those are suppressed 
compared with $M_0$.
For such suppressed values, sub-leading corrections e.g. the one-loop correction on 
the K\"ahler metric are not negligible anymore.
That introduces one-loop ambiguity into the model.
Including such corrections, at the TeV scale we obtain 
\begin{equation}\label{eq:soft-at-mirage-2}
m_{H_u}^2 \approx \delta c_{H_u}^{(\rm loop)}M^2_0, \qquad 
m_{S}^2 \approx \delta c_S^{(\rm loop)}M^2_0,
\end{equation}
with  $\delta c_{H_u}^{(\rm loop)},\delta c_S^{(\rm loop)} = {\cal
{O}}(1/8\pi^2)$.
Note that similar to Eq.(\ref{eq:soft-at-mirage-1}), 
Eq.(\ref{eq:soft-at-mirage-2}) also includes corrections of  
${\cal O}(\kappa^2M_0^2/8\pi^2)$ due to the violation of the mirage unification by $\kappa$.
That is, we obtain $m_{H_u}^2= 0, m_{S}^2= 0$ 
up to ${\cal {O}}(M_0^2/8\pi^2)$ at the TeV scale.
Similarly, at the TeV scale we can obtain,
\begin{equation}
A_\kappa = 0, 
\end{equation}
up to ${\cal {O}}(M_0/8\pi^2)$.
Because of such ambiguity, we use $A_\kappa$ 
as a free parameter, which must be small 
compared with $M_0$.
In addition, we determine the values of 
$m_{H_u}^2$, $m_S^2$ and $\mu~(=\lambda s)$ at the electroweak scale 
from the stationary conditions,  
(\ref{eq:1}), where we use the experimental value 
$m_Z = \frac{1}{\sqrt{2}} g v = 91.19$ GeV and $\tan \beta$ as a free parameter.

%%KO
Through the above procedure, the parameters, $m^2_{H_u}$, $m^2_S$ and $\mu$, 
at the electroweak scale are expressed by $\tan\beta$, $m_{H_d}^2$, $A_\lambda$ as follows, 
\begin{eqnarray}
\mu &=& \lambda \langle S \rangle = \frac{A_\lambda \tan\beta
}{2\left(1-\frac{\kappa}{\lambda}\tan\beta\right)}   \left\{1
-\sqrt{1-4 X}\right\} ,
\nonumber\\
m^2_S &=& -2\left(\frac{\kappa}{\lambda}\right)^2\mu^2
-\left(\frac{\kappa}{\lambda}\right) A_\kappa \mu
+\frac{\lambda^2}{g^2}m_Z^2
\left\{\left(\frac{A_\lambda}{\mu}+2\frac{\kappa}{\lambda}\right)\sin 2\beta -2\right\},\\
m_{H_u}^2 &=&
\frac{\tan^2\beta-1}{\tan^2\beta}\left(\frac{m_{H_d}^2}{\tan^2\beta-1}-\mu^2-\frac{m_Z^2}{2}\right), \nonumber
\label{eq:minimum}
\end{eqnarray}
where,  
\begin{equation}
X = \frac{m_{H_d}^2\left(1-\frac{\kappa}{\lambda}\tan\beta\right)}{ A_\lambda^2 \tan^2\beta}\left\{1 +\frac{\tan^2\beta}{\tan^2\beta+1}
\left(\frac{2\lambda^2}{g^2}-\frac{\tan^2\beta-1}{2\tan^2\beta}\right)\frac{m_Z^2}{m_{H_d}^2}\right\}. 
\end{equation}
%
%For $\tan\beta >> {\rm max}( 1, \lambda/\kappa)$, they are approximated as,
%KO>
For $\tan\beta \gg {\rm max}( 1, \kappa/\lambda)$, these parameters are approximated as,
%<KO
%Through the above procedure, the values of $m^2_{H_u}$, $m^2_S$ and $\mu$ 
%at the weak scale are approximately obtained as 
\begin{subequations}
\begin{align}
& \mu = \lambda \langle S \rangle \sim   \frac{m^2_{H_d}}{A_\lambda \tan \beta},  \label{eq:app-1}\\  
%& m^2_S \sim \frac12 \frac{\kappa^2\mu^2}{\lambda^2}, \label{eq:app-2}\\
& m^2_S \sim -2 \left(\frac{\kappa}{\lambda}\right)^2\left(\frac{m_{H_d}^2}{A_\lambda\tan\beta}\right)^2-\left(\frac{\kappa}{\lambda}\right)A_\kappa \left(\frac{m_{H_d}^2}{A_\lambda\tan\beta}\right)+2\frac{\lambda^2}{g^2}\frac{A_\lambda^2}{m_{H_d}^2} m_Z^2, \label{eq:app-2}\\
& m_{H_u}^2 \sim  \frac{m^2_{H_d}}{\tan^2 \beta} -
\frac{m^4_{H_d}}{A^2_\lambda \tan^2 \beta}- \frac{m^2_Z}{2} .
\label{eq:app-3}
\end{align}
\label{eq:app}
\end{subequations}
When $\tan \beta ={\cal O}(10)$, 
the values of $\mu$, $|m_{H_u}|$ and $|m_S|$ are smaller than $M_0$ 
by the factor $\tan \beta$ because $m_{H_d} \simeq A_\lambda \simeq M_0$.
Thus, the values of $\mu$ and $|m_{H_u}|$ could be of ${\cal O}(100)$ GeV 
while the other masses of the superpartners are of ${\cal O}(M_0)={\cal O}(1)$ TeV.
Then, the fine-tuning problem can be ameliorated.
%
%KO
Furthermore, one can see that the first and the second terms in the last equation  
cancel each other for our choice of $c_i$. 
The next leading contributions are of ${\cal O}(m^2_{H_d}/\tan^4\beta)$ or
${\cal O}(m^2_{H_d}\mu/\tan^2\beta A_\lambda)$. 
Thus, $m^2_Z$ is almost determined by $m^2_{H_u}$ alone and
insensitive to the value of $\mu$. 
This means that actually $\tan\beta \approx 3$ is enough to obtain
the fine-tuning of $|\partial \ln m_Z^2/\partial \ln m_{H_u}^2|^{-1}=m_Z^2/2m_{H_u}^2={\cal O}(100)$\% for $M_0\approx 1$ TeV.
In this case, $\mu$ can be as heavy as ${\cal O}(400)$ GeV without deteriorating the fine-tuning.
%KO>
The origin of this cancellation is easily understood by examining the doublet mass matrix,
\begin{equation}
{\cal L}_M = -\left( H_d, H_u^\ast\right)
{\cal M}^2_H
\left(
\begin{array}{c}
H_d^\ast \\
H_u
\end{array}
\right),
\end{equation}
where 
\begin{equation}
{\cal M}^2_H =
\left( 
\begin{array}{cc}
m_{H_d}^{2} +\mu^2 & -A_\lambda \mu \\
-A_\lambda \mu & m_{H_u}^2+\mu^2
\end{array}
\right)
\approx
\left( 
\begin{array}{cc}
M_0^2 +\mu^2 & -M_0 \mu \\
-M_0 \mu & \mu^2
\end{array}
\right).
\end{equation}
The modulus mediated contribution $M_0$ cancels in the determinant of the mass matrix, 
${\rm det}({\cal M}^2_H) \approx \mu^4$. 
The heavy mode has mass of ${\cal O}(M_0)$, then the mass of 
the light mode is suppressed as $\mu^2/M_0\approx \mu/\tan\beta$ and a flat direction appears 
along $H_u/H_d \approx M_0/\mu \approx \tan\beta$.
%KO<
This mechanism was previously observed in \cite{Choi:2006xb}
in the context of the MSSM.
In the NMSSM, the relation $m_{H_d} \approx A_\lambda$
is well controlled up to the leading contribution of the modulus mediation, in contrast to the $B$-term (in place of $A_\lambda$) in the MSSM, which is a remnant of
the fine-tuned cancellation between the terms of ${\cal O}(m_{3/2})$
and subject to uncontrolled corrections.

In the following section, we show numerically the spectrum 
of our model.

\subsection{Spectrum}

Here, we study numerically the spectrum of our model.
Before showing numerical results, we recall our parameters.
In our analysis, the free parameters are 
$\lambda$, $\kappa$, $\tan \beta$ and $A_\kappa$ given at the SUSY scale  
and $M_0$ given at $M_{GUT}$. As the SUSY scale, we choose $M_{SUSY} = \sqrt{m_{\tilde{t}_1}m_{\tilde{t}_2}}\simeq M_0/\sqrt{2}$.
Using them, we determine all of the soft SUSY breaking 
parameters except for $m_{H_u}^2$, $m_S^2$ and $\mu$,
which are determined by using the stationary condition of the Higgs potential. 
Note that in the following numerical analysis we include corrections
from the 1-loop effective potential,
\begin{equation}
V_{\text{1-loop}} = \frac{1}{64\pi^2} Str
\left[{\cal M}^4\left\{\ln\left(\frac{{\cal M}^2}{M_{SUSY}^2}\right)-\frac{3}{2}\right\} \right],
\label{eq:1-loop-corr}
\end{equation}
in the stationary conditions (\ref{eq:1}), where $\cal M$ represents the mass matrix of our model and $Str$ denotes the supertrace.
In all of the following numerical analysis, 
we use $A_\kappa = - 100$ GeV as a typical value of $A_\kappa$.
When $\lambda$ and/or $\kappa$ are large at the electroweak scale, 
they blow up below the GUT scale.
Thus, we have constrains on large values of 
$\lambda$ and $\kappa$ 
by requiring that those do not blow up below 
the GUT scale. 

Figure $1$ shows the lightest CP-even Higgs mass $m_{h1}$, 
soft scalar masses of $H_u$ and $S$, $m_{H_u}$ and $m_{S}$, 
and $\mu$  
for $M_0=1200$ GeV and $\tan \beta =3$, in panels (a), (c), (d) and (e), respectively.
The panel (b) in the figure shows the coupling squared 
between the lightest CP-even Higgs and 
the Z bosons, $g^2_{ZZh1}$, as the ratio to the one in the SM, 
i.e. $g^2_{ZZh1}/g^2_{\rm SM}$. 
The second lightest CP-even Higgs mass $m_{h2}$ is also plotted in panel (f).

In the figure, the red curve corresponds to the values of 
$\lambda$ and $\kappa$ at the electroweak scale, which 
blow up at the GUT scale.
Thus, we exclude the outside of this curve.
The gray region around $\kappa=0$ corresponds to the region where 
the tachyonic mode appears in the Higgs sector.
The yellow region is excluded because   
the Higgs potential has the false vacua studied 
in Ref. \cite{Kanehata:2011ei}, 
deeper than the realistic vacuum.
{}From these constraints, the region with small $\kappa/\lambda$ is 
disfavored.
The gray region around the red curve indicates the region where the tree level
Higgs mass becomes tachyonic and the iterative procedure we employed does
not work to estimate the stationary conditions. 
The quantum corrections \eqref{eq:1-loop-corr} could lift the tachyonic mass,
however, we do not calculate it because the region
has already been excluded by the LEPII bound ($m_{h_1} > 114.4$ GeV).

The value of $\mu$ is around $200-400$ GeV, 
which is consistent with the rough estimation in 
Eq.~(\ref{eq:app-1}), i.e.
$\mu \sim m_{H_d}^2/(A_\lambda \tan \beta) \sim M_0/\tan \beta$.
%%KO>
Obviously 
the expansion in Eq.~(\ref{eq:app-1}) becomes worse for $\kappa/\lambda \gtrsim \tan\beta$,
%while for large $\lambda$, $m_{H_d}$ and $A_\lambda$ deviate from $M_0$ due to the running between the mirage scale to the electroweak scale. Then the approximation is not good in these regions.
while the expansion holds well for $\kappa/\lambda \lesssim 0.3$ where 
the value of $|m_{H_u}|$ is around $100 - 200$ GeV. 
In this region the fine-tuning of the parameters for the electroweak symmetry breaking
is of ${\cal O}(10) \%$, even though most of the superpartners have heavy masses of ${\cal O}(M_0)$.
A large value of $\mu\simeq 400$ GeV potentially degrades the fine-tuning, however
the cancellation renders $\mu$ irrelevant to $m_Z$ at the leading order of the expansion by $1/\tan\beta$. 
%The mechanism works well for $\kappa/\lambda \sim 0.1$.
%In all the surviving parameter space, both mass scales of $\mu$ and 
%$|m_{H_u}|$ 
%are around 200 GeV and the fine tuning is of
%${\cal O}(10) \%$.
%%%<KO
%Thus, the fine-tuning is improved even though most of 
%superpartners have heavy masses of ${\cal O}(M_0)$.

%%KO>
The value of $|m_S|$ is roughly estimated as $|m_S| \sim (\sqrt 2 \kappa/ \lambda) \mu$
in Eq.~(\ref{eq:app-2}). Thus, it is found that $|m_S| \approx 400$ GeV for $\kappa \approx \lambda$ 
and $|m_S|$ increases (decreases) as $\kappa/\lambda $ increases (decreases).
%%<KO
The value of $|m_S|$ is expected to be suppressed in the 
TeV scale mirage scenario.
However, a large value of $\kappa/\lambda$ leads large $|m_S|$ through the 
stationary condition like $|m_S| \sim 500$ GeV. 
Such a large value would not be realized in our TeV scale mirage
mediation scenario, because $m^2_S$ must be suppressed compared with $M_0^2$.
Thus, the region with large $\kappa/\lambda$ and $|m_S| > {\cal O}(M_0/\sqrt{8\pi^2})$ 
is disfavored. In the panel (d), the region with $|m_S| > M_0/\sqrt{8\pi^2}$ is filled in pink. 
It is interesting that the region favored by the TeV scale mirage mediation
exactly corresponds to the region where the fine-tuning is ameliorated.
Note that the condition (\ref{eq:mir-condition}) also holds well for this region.

%%KO>
In Fig.~1.(a), the lightest CP-even Higgs boson, which dominantly consists of the doublet scalar here, has a 
mass $m_{h_1}$ around $80$--$125$ GeV.
We estimated $m_{h_1}$ using NMHDECAY in the NMSSMTools package \cite{Ellwanger:2005dv}.
We calculated the minimum of the effective potential renormalized
at $M_{SUSY}$ by the iteration starting
from the tree-level minimum. Then we used the resultant $\mu$ as 
an input of the NMSSMTools. The pole mass $m_t = 172.9$ GeV was used 
in our calculation.
%by the effective Lagrangian approach described in the Appendix B.
%We evaluate the Higgs potential at $m_t$ and use the pole mass 
%$m_t = 172.9$ GeV. Alternatively,
%the evaluation at $m_Z$ generally pushes $m_{h_1}$ by
%$\sim 5$ GeV. Thus the ambiguity of this size should be understood.
%%The heavier Higgs mass $m_{h_1}$ appear in the region with small 
%%$\kappa/\lambda$ outside of the excluded green region. 
The qualitative behavior of the mass of the SM-like Higgs boson is given by
\begin{eqnarray}
m_{\mathrm{SM}}^2 &\simeq& m_Z^2 \cos^22 \beta +\lambda^2v^2\sin^2 2 \beta
-\frac{\lambda^2}{\kappa^2}v^2
(\lambda -\kappa \sin 2\beta)^2 \nonumber\\
&& +\frac{3m_t^4}{4\pi^2v^2}
\left(\ln\left(\frac{m_{\tilde{t}}^2}{m_t^2}\right)
+\frac{A_t^2}{m_{\tilde{t}}^2}
\left(1-\frac{A_t^2}{12m_{\tilde{t}}^2}\right)
\right),
\end{eqnarray}
for $\kappa s \gg |A_\kappa|, |A_\lambda|$ in \cite{Ellwanger:2009dp}.
The first term and the fourth term are the tree-level contribution and the radiative correction in the MSSM, respectively.
The second term comes from the new quartic Higgs couplings in the NMSSM, 
and the third term comes from the mixing of the doublet scalars with the singlet. 
The third term is always negative because the mixing reduces the lightest eigenvalue of the mass matrix.
Although the above approximation does not apply in the figure, the behavior of the higgs mass can be understood 
in terms of the mixing. The effect of the mixing undermines that of  
the additional quartic coupling $\lambda^2\sin^2 2\beta$ and suppresses the Higgs mass
except for the narrow region $4 \lesssim \lambda/\kappa \lesssim 8$.
The LHC observation, $m_{h_1} \approx 125$ GeV, is satisfied only around $(\lambda,\kappa)=(0.7,0.11)$ where the mixing vanishes.
Such a region may be realized as a quasi-infrared fixed point
if $\lambda$ has a strong dynamics origin at the GUT scale, while
$\kappa$ is suppressed due to the approximate Peccei-Quinn symmetry.
It is important to stress again that this region is favored by the TeV scale mirage mediation and the fine-tuning.
%KO>
The mixing of the doublet and the singlet Higgs scalars in the Lagrangian is obtained by rotating the doublet Higgs 
by $\beta~(=\tan^{-1} v_u/v_d)$
\begin{align}
\Delta{\cal L} = -2v \lambda \left[\mu-\left(A_\lambda +2\kappa s \right) \cos\beta\sin\beta\right] h \Delta S_r, 
\end{align}
where $h = \mathrm{Re}(H_d) \cos\beta +\mathrm{Re}(H_u)\sin\beta -v$ and $\Delta S_r = \mathrm{Re}(S)-s$ denote 
dynamical degree of freedom of the corresponding Higgs fields.
Since our TeV scale mirage mediation scenario leads 
$\mu \approx M_0/\tan\beta$ and $A_\lambda \approx M_0$, 
the above term is approximated as,
\begin{align}
\Delta{\cal L} = 4v\frac{\mu}{\tan\beta} \left[ \frac{\kappa}{\lambda} +{\it O}\left(\frac{1}{\tan^2\beta}\right) \right] h \Delta S_r.
\end{align}
%<KO
Thus the mixing is automatically suppressed by $(\kappa/\lambda)/\tan\beta$ in our scenario.

%Then $m_{h_1}$ decreases for small $\kappa/\lambda$.
%%<KO
%
The coupling of the lightest CP-even Higgs boson 
to the $Z$ boson, $g_{ZZh_1}$, is almost the same as 
the one in the standard model in most of the parameter space
%KO>
 as expected.
%KO<
We show the ratio  $g^2_{ZZh_1}/g^2_{\rm SM}$ in the figure, where $g_{\rm SM}$ denotes the Higgs coupling to 
the Z boson in the standard model.
The mixing between the doublet and the singlet is minimized around $m_{h_1} \approx 125$ GeV, where $g_{ZZh_1}$ also approaches to its SM value.

Table 1 shows examples of spectra and $g^2_{ZZh_1}/g^2_{\rm SM}$
for $(\lambda,\kappa)=(0.10,0.40)$, $(0.40,0.10)$ and $(0.70,0.11)$.
In the table, $m_{hi}$ for $i=1,2,3$ and $m_{ai}$ for $i=1,2$ 
denote three CP-even Higgs masses and two CP-odd Higgs masses, 
respectively.
Also, $m_{\tilde t_{1,2}}$ denote two eigenvalues of 
stop masses.
Other squark and slepton masses depend on their values of $c_i$, 
and they are of ${\cal O}(M_0)$ unless $c_i =0$. 
The second lightest CP-even Higgs boson $h_2$  is lighter 
for  $(\lambda,\kappa)=(0.40,0.10)$ and  $(0.70,0.11)$ than 
the one for $(\lambda,\kappa)=(0.10,0.40)$.
Its dominant component is the singlet Higgs boson $S$.
Its mass decreases as $\kappa/\lambda$ decreases.
Then, the above behavior of $m_{h_2}$ occurs.
On the other hand, the dominant component of the 
heaviest CP-even Higgs boson $h_3$  is the down-type Higgs boson $H_d$. 
Its mass is heavy and almost equal to $M_0$, 
%%KO
%except for large $\lambda$ where the running between the mirage scale and the electroweak scale is sizable.
independent of $\lambda$ and $\kappa$.
%%KO
By the same reason, the mass of 
the heavier CP-odd Higgs boson $a_2$ is almost the same as 
$m_{h_3}$ as well as $M_0$.
%The lightest CP-odd Higgs boson $a_1$ is lighter
%and has a mass similar to $m_{h_2}$ for small $\kappa$.
%KO>
The lightest CP-odd Higgs boson $a_1$ is lighter for small $\kappa$
 due to the approximate Peccei-Quinn symmetry.
%<KO
Also, three gaugino masses are almost the same as $M_0$.
%KO>
It might be challenging but interesting subject
 to observe these light extra-Higgs bosons
 through the small mixing with the doublets in LHC and ILC.
%<KO

\begin{figure}[h]
\begin{center}
\begin{tabular}{l @{\hspace{10mm}} r }
\includegraphics[height=55mm]{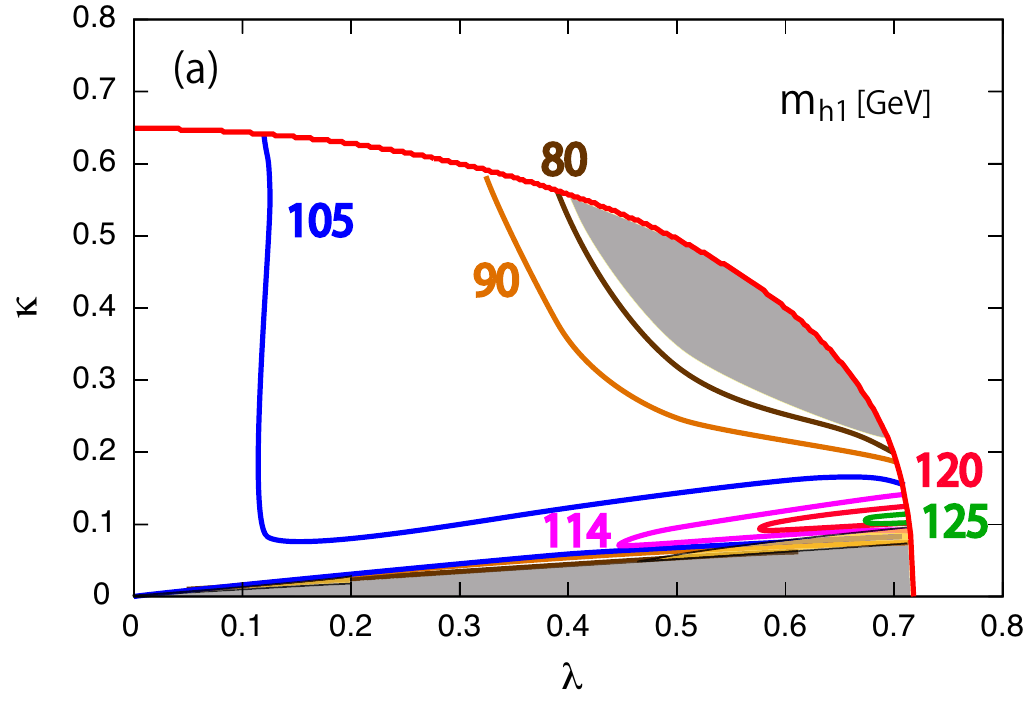} &
\includegraphics[height=55mm]{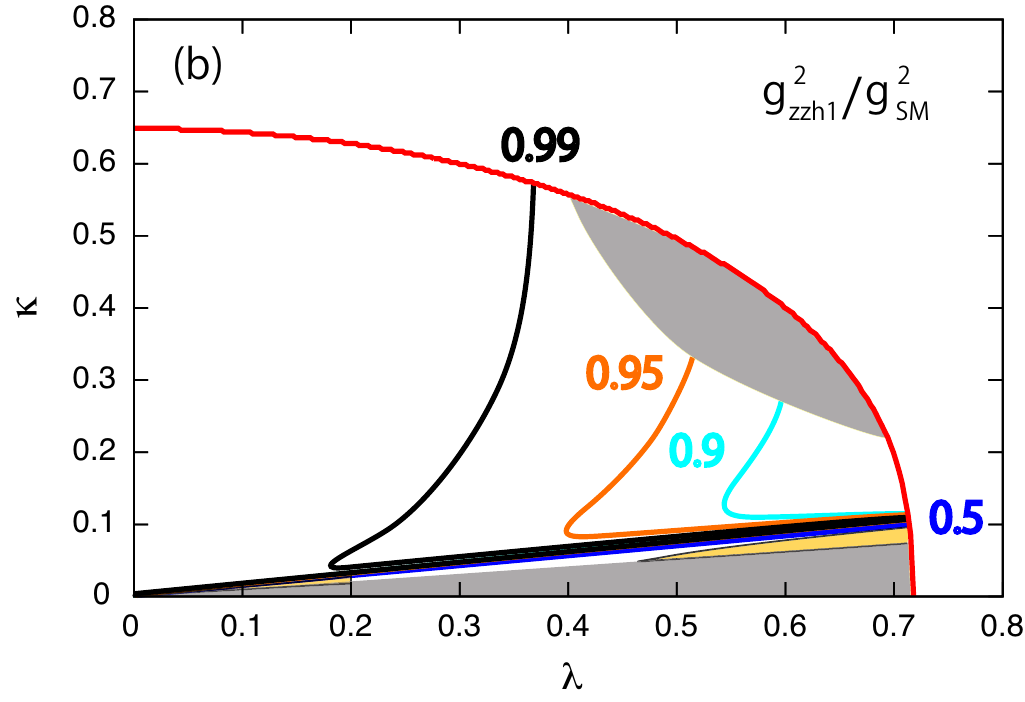} \\
\rule{0cm}{10mm} & \\
\includegraphics[height=55mm]{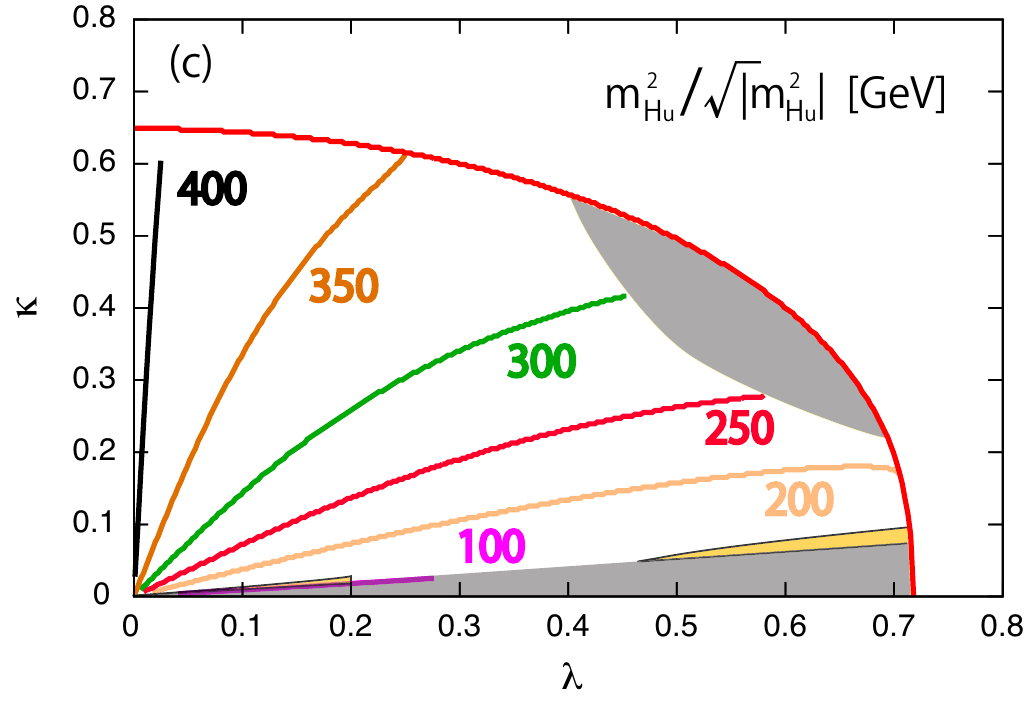} &
\includegraphics[height=55mm]{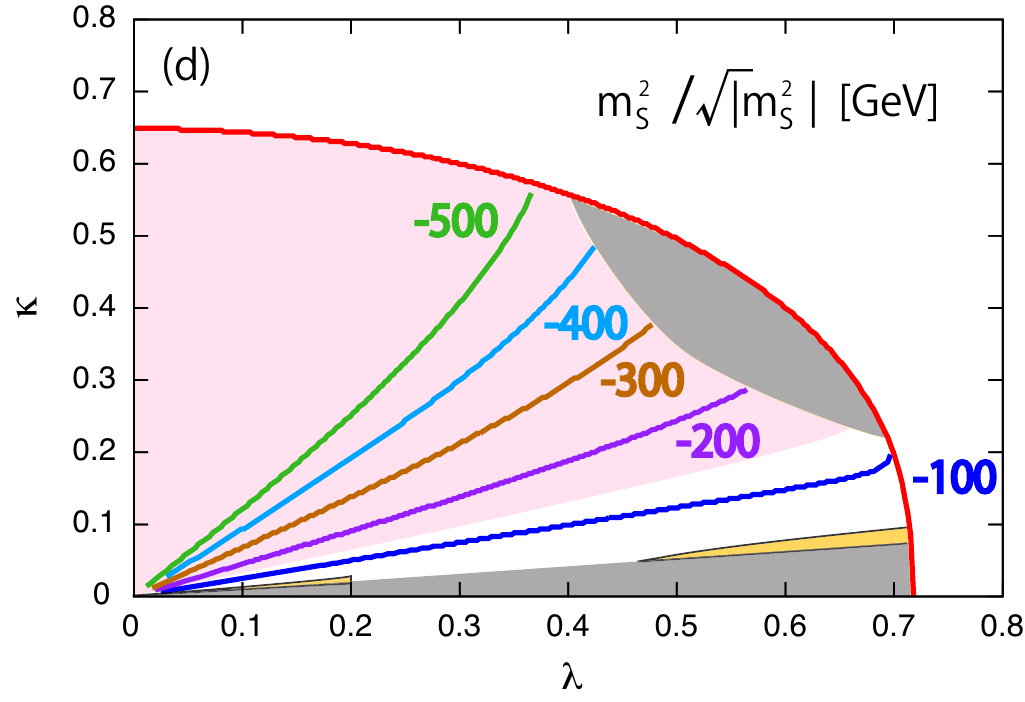} \\
\rule{0cm}{10mm} & \\
\includegraphics[height=55mm]{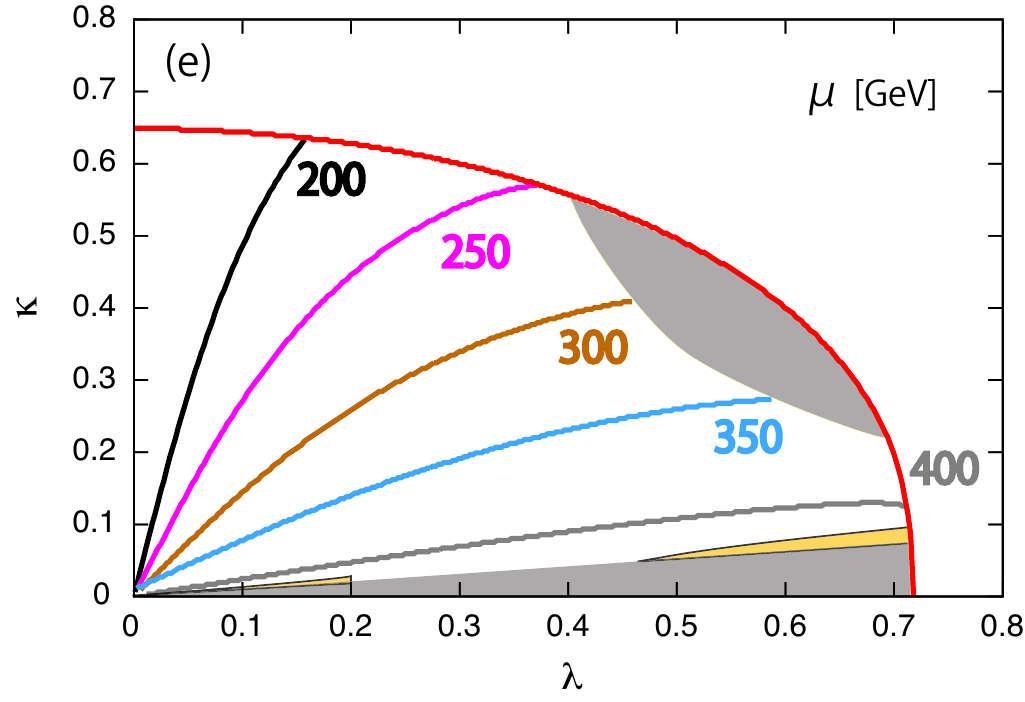} & 
\includegraphics[height=55mm]{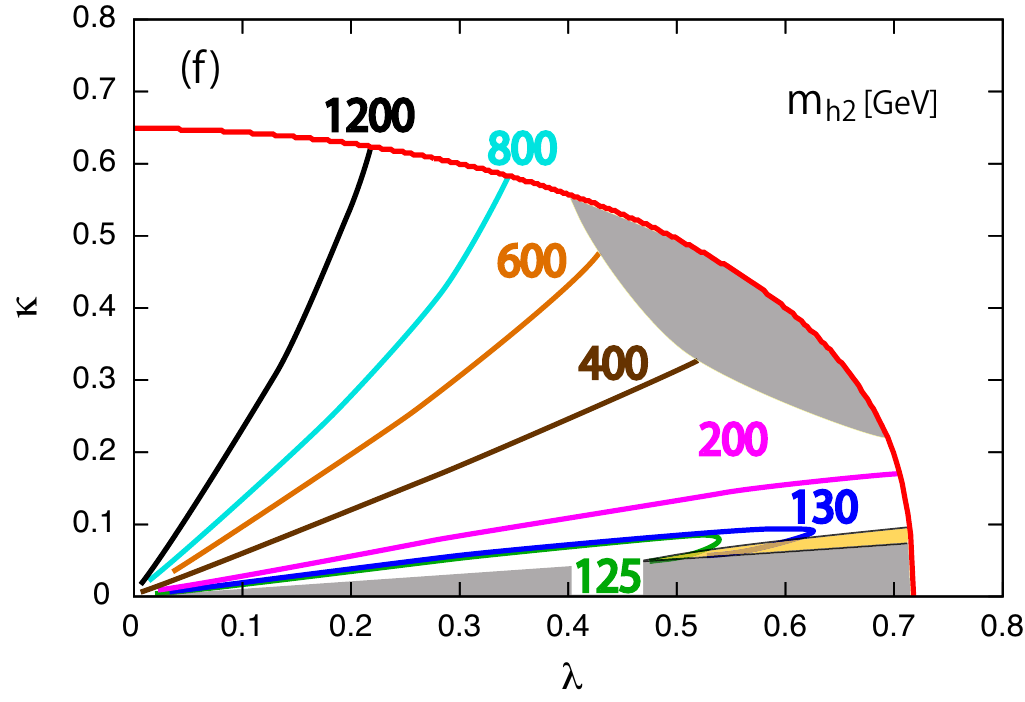}
\end{tabular}
\caption{Constraints and SUSY breaking parameters.
We use $M_0 = 1200GeV, \alpha=2, c_{Hd}=1, c_{Hu}=0, c_S=0, c_{\tilde{Q}}=0.5, c_{\tilde{t}}=0.5$
and $\tan\beta=3 , A_\kappa=-100\mathrm{GeV}$.
For detail of the shaded regions, see the text.
The TeV scale mirage mediation disfavors the region $|m_S|\gtrsim M_0/\sqrt{8\pi^2}\sim 100$ GeV (pink shaded).
}
\end{center}
\end{figure}

\begin{table}[htbp]
\begin{center}
\begin{tabular}{|c|c|c|c|c|}
\hline
$(\lambda,\kappa)$ & $(0.10,0.40)$ &  $(0.40,0.10)$ & $(0.70,0.11)$\\ \hline \hline
$m_{h1}$ & 105 GeV & 107 GeV & 126 GeV \\ \hline
$m_{h2}$ & 1261 GeV & 182 GeV & 138 GeV \\ \hline
$m_{h3}$ & 1700 GeV & 1312 GeV & 1320 GeV \\ \hline
$m_{a1}$ & 512 GeV & 181 GeV & 158 GeV \\ \hline
$m_{a2}$ & 1260 GeV & 1311 GeV & 1321 GeV\\ \hline
$g_{ZZh_1}^2/g^2_{\rm SM}$ & 1.00 & 0.95 & 0.94 \\ \hline

$m_{Hd}^2$ & $1.39 \times 10^6 \mathrm{GeV}^2$ 
        & $1.39 \times 10^6 \mathrm{GeV}^2$ 
        & $1.40 \times 10^6 \mathrm{GeV}^2$ \\ \hline
$m_{Hu}^2$ & $1.29 \times 10^5 \mathrm{GeV}^2$ 
        & $5.58 \times 10^4 \mathrm{GeV}^2$ 
        &  $3.06 \times 10^4 \mathrm{GeV}^2$ \\ \hline
$m_{S}^2$ & $-1.42 \times 10^6 \mathrm{GeV}^2$ 
        & $-1.02 \times 10^4 \mathrm{GeV}^2$ 
        & $-4.93 \times 10^3 \mathrm{GeV}^2$ \\ \hline
$m_{\tilde{t}_{1}}$ & 823 GeV & 823 GeV  &  823 GeV \\ \hline
$m_{\tilde{t}_{2}}$ & 849 GeV & 849 GeV &  849 GeV \\ \hline
%$m_{\tilde{b}_{1}}$ & 900.4 GeV & \\ \hline
%$m_{\tilde{b}_{2}}$ & 1007.3 GeV & \\ \hline
$\mu$ & 217 GeV &  396 GeV & 407 GeV \\ \hline
\end{tabular}
\caption{Spectra for $(\lambda, \kappa)=(0.10,0.40)$, $(0.4,0.10)$ and $(0.7,0.11)$ 
with $\tan\beta=3$ and $M_0 = 1200$GeV.}
\end{center}
\end{table}

\begin{table}[htbp]
\begin{center}
\begin{tabular}{|c|c|c|c|c|}
\hline
$(\lambda,\kappa)$ & $(0.10,0.40)$ &  $(0.40,0.10)$ & $(0.70,0.11)$\\ \hline \hline
$m_{h1}$ & 107 GeV & 110 GeV & 128 GeV \\ \hline
$m_{h2}$ & 1574 GeV & 231 GeV & 172 GeV \\ \hline
$m_{h3}$ & 2134 GeV & 1637 GeV & 1641 GeV \\ \hline
$m_{a1}$ & 572 GeV & 200 GeV & 172 GeV \\ \hline
$m_{a2}$ & 1573 GeV & 1636 GeV & 1645 GeV \\ \hline
$g_{ZZh_1}^2/g^2_{\rm SM}$ & 1.00 & 0.98 & 0.99 \\ \hline

$m_{Hd}^2$ & $2.25 \times 10^6 \mathrm{GeV}^2$
        & $2.18 \times 10^6 \mathrm{GeV}^2$
        & $2.18 \times 10^6 \mathrm{GeV}^2$ \\ \hline
$m_{Hu}^2$ & $2.03 \times 10^5 \mathrm{GeV}^2$
        & $5.02 \times 10^4 \mathrm{GeV}^2$
        & $3.84 \times 10^4 \mathrm{GeV}^2$ \\ \hline
$m_{S}^2$ & $-2.25 \times 10^6 \mathrm{GeV}^2$
       & $-1.92 \times 10^4 \mathrm{GeV}^2$
       & $-9.38 \times 10^3 \mathrm{GeV}^2$ \\ \hline
$m_{\tilde{t}_{1}}$ &1023 GeV & 1023 GeV & 1023 GeV \\ \hline
$m_{\tilde{t}_{2}}$ & 1055 GeV & 1055 GeV & 1055 \\ \hline
%$m_{\tilde{b}_{1}}$ & 1137.9 GeV \\ \hline
%$m_{\tilde{b}_{2}}$ & 1277.3 GeV \\ \hline
$\mu$ & 272 GeV & 495 GeV & 508 GeV \\ \hline
\end{tabular}
\caption{Spectra for $(\lambda, \kappa)=(0.10,0.40)$, $(0.40,0.10)$ and $(0.70,0.11)$
with $\tan\beta=3$ and $M_0 = 1500$GeV.}
\end{center}
\end{table}

Figure 2 shows the same as Fig. 1 except for $M_0 = 1500$ GeV.
Most of mass parameters become larger than 
those for $M_0=1200$ GeV.
%The favored region by the TeV scale mirage mediation becomes narrower, while
%However, the value of $\mu$ is still around $100-300 $ GeV, 
%and the value of $|m_{H_u}|$ is also around $100 - 200$ GeV.
%Then, the required fine-tuning is of ${\cal O}(10\%)$.
%Also,  $|m_S|$ becomes larger.
%%and the disfavored region with large  $|m_S|$  such as 
%%$|m_S| \sim 400 - 500$ GeV becomes wider.
%KO>
The lightest CP-even Higgs mass becomes heavy due to the heavier stop and
the portion of the region with $m_{h_1} \approx 125$ GeV increases.
%<KO
Table 2 is the same as Table 1 except $M_0 = 1500$ GeV. 
Obviously, all of masses become heavier than those in 
Table 1.
The behavior of $m_{h_2}$, $m_{h_3}$ and $m_{a_2}$ 
is the same as the one in Table 1.
%KO>
%If we further increase $M_0$, most of spectrum increase and the favored region
%becomes narrower in the $(\lambda,\kappa)$ plane. 
%<KO
For completeness, we also plot the figure for $M_0=1700$ GeV in Fig. 3 and list the spectra and the coupling for $M_0=1700$ GeV in Table. 3.

\begin{figure}[h]
\begin{center}
\begin{tabular}{l @{\hspace{10mm}} r }
\includegraphics[height=55mm]{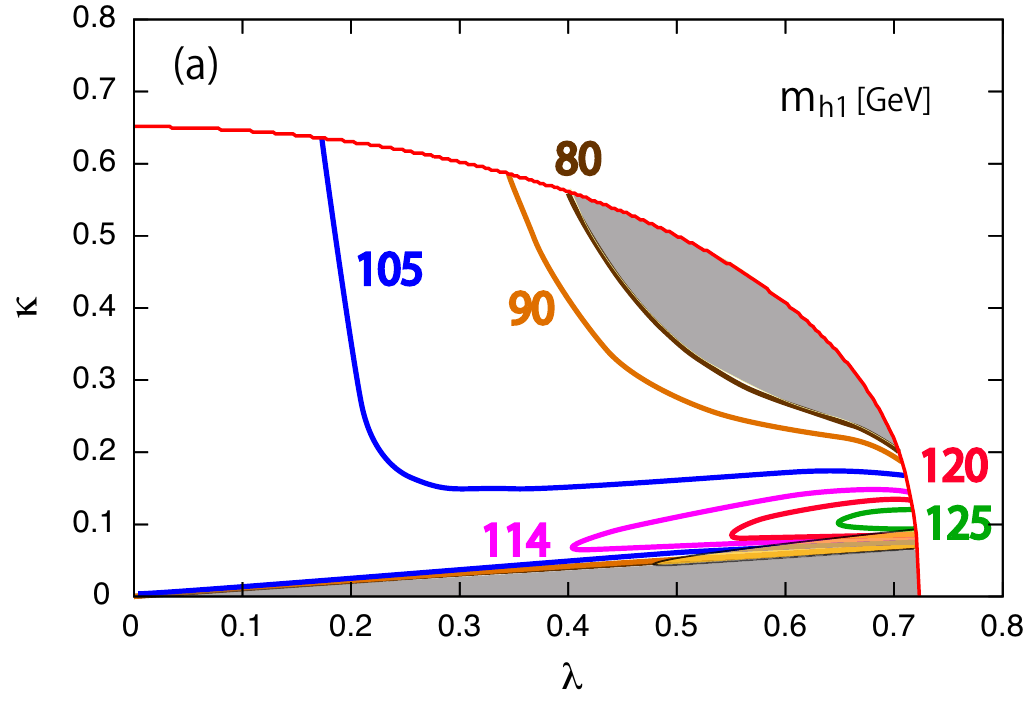} &
\includegraphics[height=55mm]{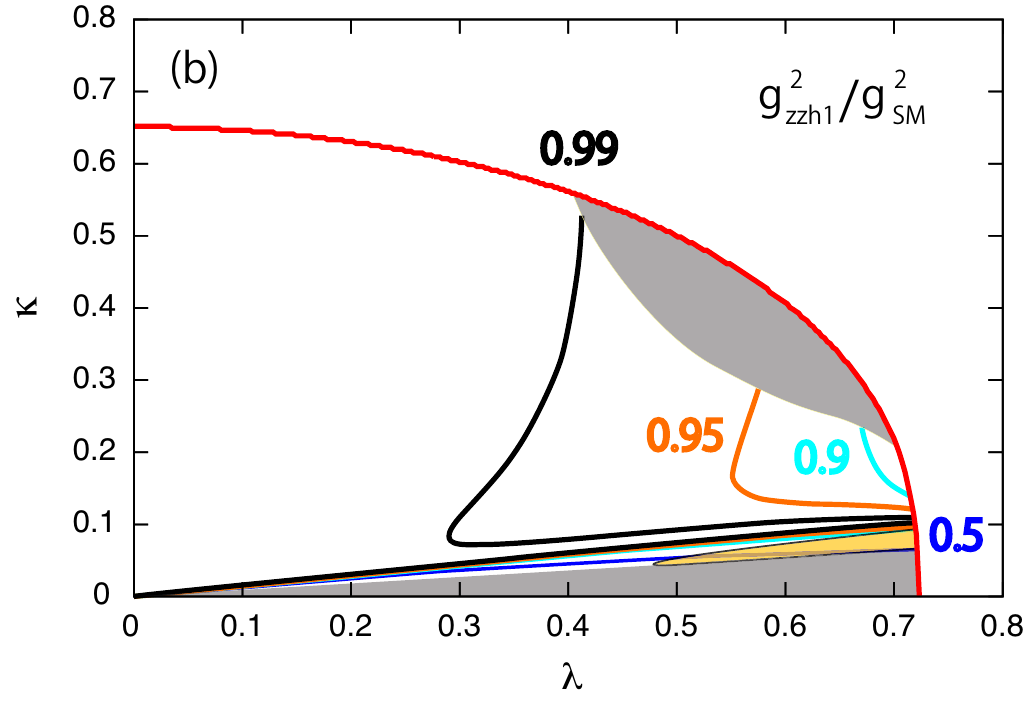} \\
\rule{0cm}{10mm} & \\
\includegraphics[height=55mm]{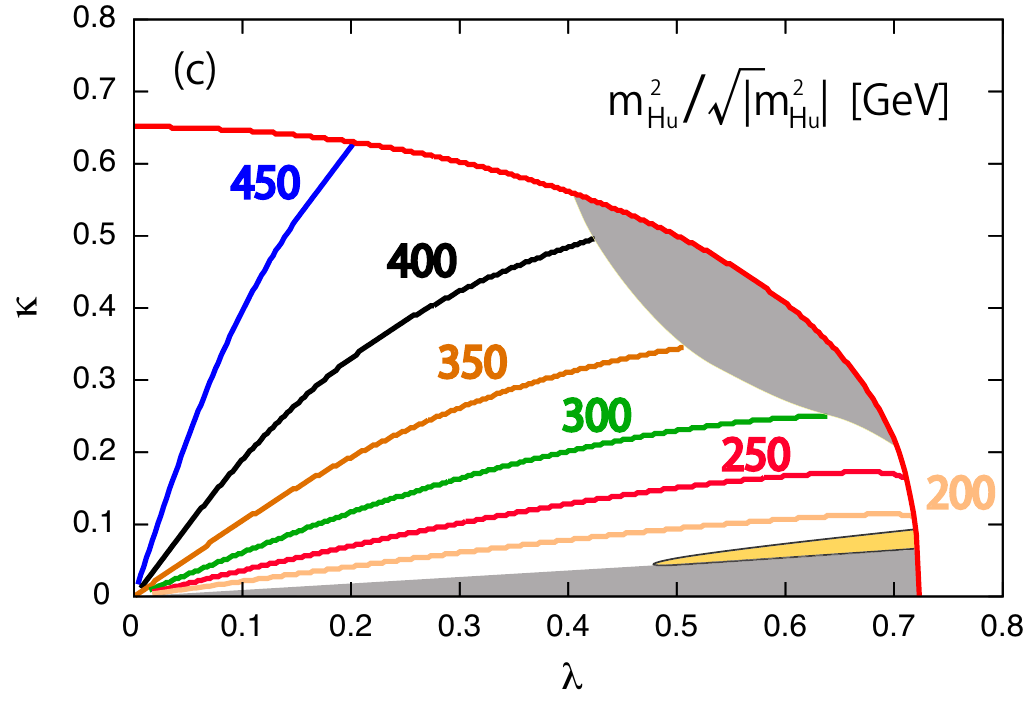} &
\includegraphics[height=55mm]{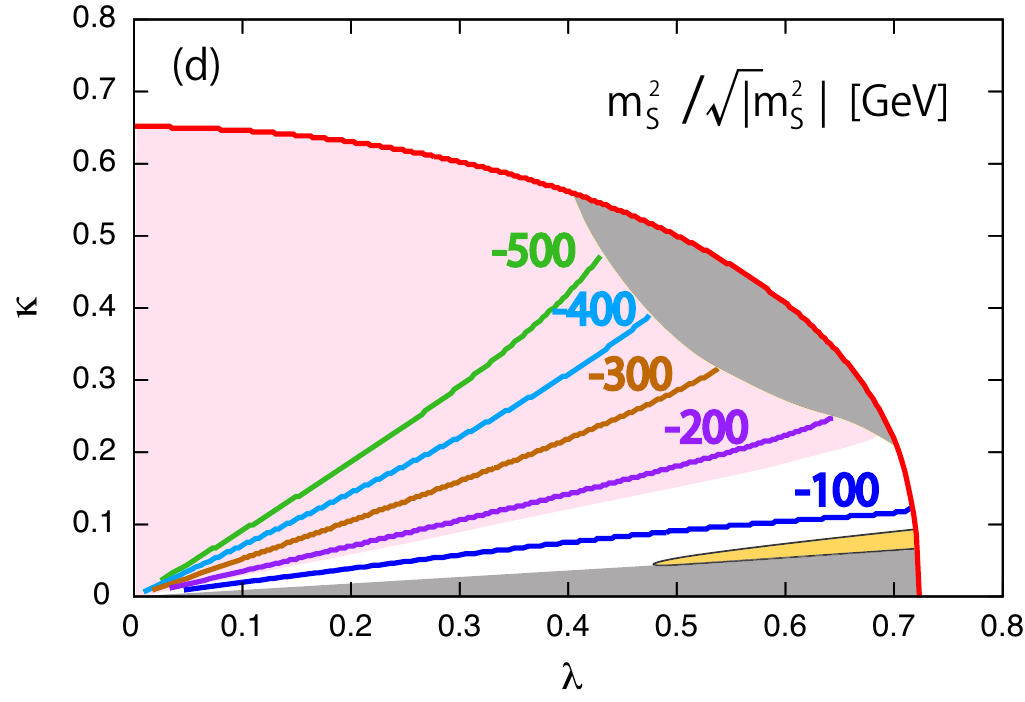} \\
\rule{0cm}{10mm} & \\
\includegraphics[height=55mm]{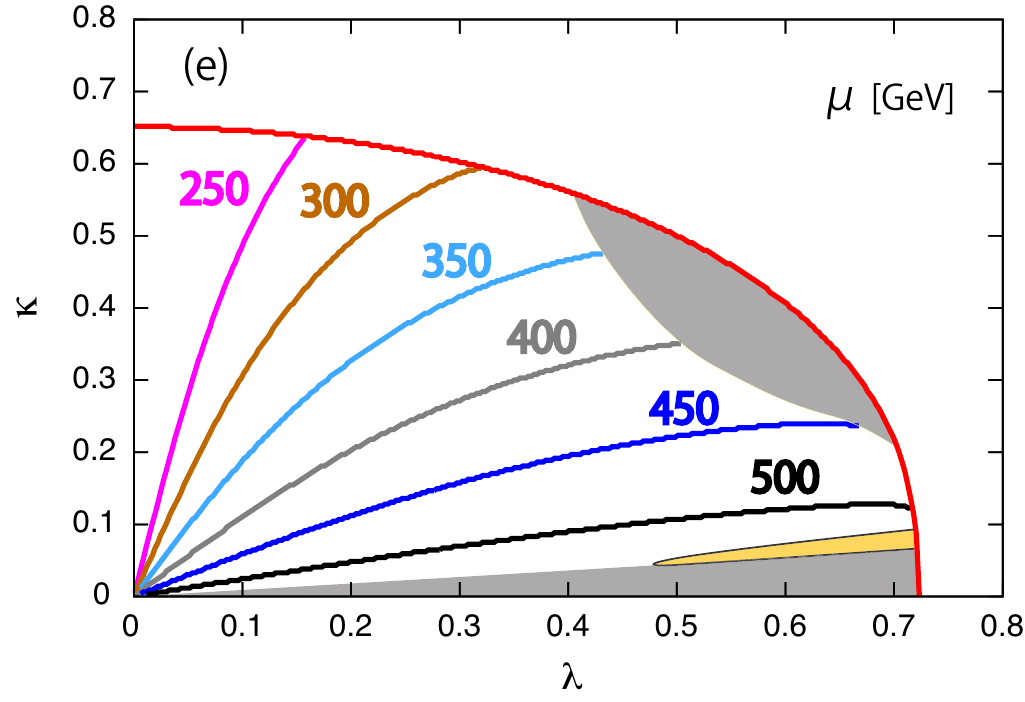} & 
\includegraphics[height=55mm]{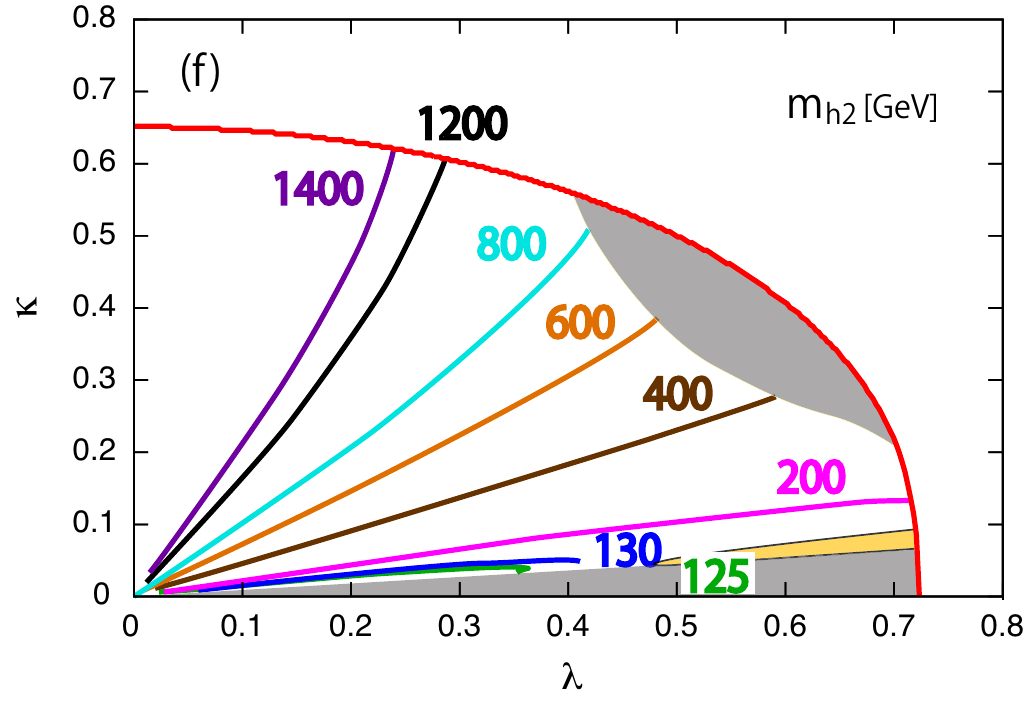}
\end{tabular}
\caption{Constraints and SUSY breaking parameters.
We use $M_0 = 1500GeV, \alpha=2, c_{Hd}=1, c_{Hu}=0, c_S=0, c_{\tilde{Q}}=0.5, c_{\tilde{t}}=0.5$
and $\tan\beta=3 , A_\kappa=-100\mathrm{GeV}$.}
\end{center}
\end{figure}

%If we further increase $M_0$, most of spectrum increase.
%For example, for  $M_0=1700$ GeV, we obtain 
%$|\mu|$ around $300 - 600$ GeV 
%and $|m_{H_u}|$ around $500 - 200$ GeV as shown in Fig. 3.
%The favored region becomes narrower but still exists
% and we can obtain the larger potion of the region
% with $m_{h_1} \approx 125$ GeV.
%We also show the examples of spectra and $g^2_{ZZh_1}/g^2_{\rm SM}$ in table 3.
%%Then, the fine-tuning is still mild, but we can 
%%obtain the heavier mass of the lightest Higgs boson.

\begin{table}[h]
%\begin{table}[htbp]
\begin{center}
\begin{tabular}{|c|c|c|c|c|}
\hline
$(\lambda,\kappa)$ & $(0.10,0.40)$ &  $(0.40,0.10)$ & $(0.70,0.11)$\\ \hline \hline
$m_{h1}$ & 108 GeV & 112 GeV & 128 GeV \\ \hline
$m_{h2}$ & 1782 GeV & 264 GeV & 196 GeV \\ \hline
$m_{h3}$ & 2422 GeV & 1853 GeV & 1861 GeV \\ \hline
$m_{a1}$ & 609 GeV & 202 GeV & 181 GeV \\ \hline
$m_{a2}$ & 1782 GeV & 1854 GeV & 1861 GeV \\ \hline
$g_{ZZh_1}^2/g^2_{\rm SM}$ &  1.00 & 0.99 & 1.00 \\ \hline

$m_{Hd}^2$ & $2.79 \times 10^6 \mathrm{GeV}^2$
        & $2.80 \times 10^6 \mathrm{GeV}^2$
        & $2.80 \times 10^6 \mathrm{GeV}^2$ \\ \hline
$m_{Hu}^2$ & $2.62 \times 10^5 \mathrm{GeV}^2$ 
        & $6.56 \times 10^4 \mathrm{GeV}^2$
        & $5.09 \times 10^4 \mathrm{GeV}^2$ \\ \hline
$m_{S}^2$ & $-2.91 \times 10^6 \mathrm{GeV}^2$
       & $-2.66 \times 10^4 \mathrm{GeV}^2$
       & $-1.31 \times 10^4 \mathrm{GeV}^2$ \\ \hline
$m_{\tilde{t}_{1}}$ & 1157 GeV & 1157 GeV & 1157 GeV\\ \hline
$m_{\tilde{t}_{2}}$ & 1193 GeV & 1193 GeV & 1193 GeV\\ \hline
%$m_{\tilde{b}_{1}}$ & 1137.9 GeV \\ \hline
%$m_{\tilde{b}_{2}}$ & 1277.3 GeV \\ \hline
$\mu$ & 308 GeV & 560 GeV & 575 GeV \\ \hline
\end{tabular}
\caption{Spectra for $(\lambda, \kappa)=(0.10,0.40)$, $(0.40,0.10)$ and $(0.70,0.11)$
with $\tan\beta=3$ and $M_0 = 1700$GeV.}
\end{center}
\end{table}

Figure 4 shows the same as Fig. 1 except for $\tan\beta=5$.
The approximation in Eq.~(\ref{eq:app-2}) and the cancellation
work well for $\kappa/\lambda \lesssim 5$.
The up-type Higgs mass $|m_{H_u}|$ is around $100-250$ GeV in most of the parameter space,
while the region, $\kappa/\lambda \lesssim 0.6$, satisfies $|m_S|\lesssim 100$ GeV
and is favored by the TeV scale mirage mediation.
%KO>
In this region, the singlet becomes lighter than the doublet and  
%The region where the mixing between the doublets and the singlet vanishes
%disappears and the lightest CP-even Higgs mass can not reach $125$ GeV due to the mixing.
%However, 
 $g^2_{ZZh1}/g_{SM}^2$ can decrease to ${\it O}(10)$ \% near the border of the excluded region by the false vacuum.
In such a region the lightest CP-even Higgs boson ($m_{h_1}\simeq 80-90$ GeV) 
could escape the LEPII bound and the second lightest CP-even Higgs boson $m_{h_2} \approx 125$ GeV gives the signal observed in LHC \cite{Belanger:2012tt}. Note that the small mixing with the singlet scalar enhances the second lightest Higgs mass in contrast to 
the lightest Higgs case and makes it easier to obtain the LHC value.
%<KO
%
%The signal strength of $h_2 \to ZZ$ is reduced by ${\it O}(10)$ \% relative to the SM because $g^2_{ZZh1}+g^2_{ZZh2} \approx g_{SM}^2$ if we decouple $h_3$. 
%The mixing also affects the production and leads further suppression. Similar reduction of the signal strength due to the mixing applies to the other channels, which will be confirmed or excluded by the future measurement of the Higgs coupling in LHC and ILC.
%KO>
The production cross section of $h_2$ ( mainly composed of $H_u$ ) through the gluon fusion or the vector boson fusion is reduced by ${\it O}(10)$ \% relative to the SM due to the mixing with the singlet. 
The branching ratio $BR(h_2 \to ZZ, WW, \gamma\gamma)$ is also reduced.
This is because the width $\Gamma(h_2 \to b\bar{b})$ does not change due to the suppressed mixing between the heavy $H_d$ and the singlet, while the widths $(h_2 \to ZZ, WW, \gamma\gamma)$ are reduced due to the mixing between $H_u$ and the singlet.
Such a reduction will be confirmed or excluded by the future measurement of the Higgs coupling in LHC and ILC.
%<KO
Table 4 shows the examples of the mass spectra and $g^2_{ZZh_1}/g^2_{\rm SM}$ for $(\lambda,\kappa)=(0.10,0.40)$, $(0.40,0.10)$ and $(0.70,0.11)$.

%In this region the cancellation between $\mu$ and $m_{H_d}/\tan\beta$ 
%ensures $|m_{H_u}| \lesssim 200 $ GeV and the fine-tuning of ${\cal O}(10)$ \%,
%even for large value of $\mu \lesssim 600$ GeV.
%The cancellation degrades for $\lambda \gtrsim 0.6$ because of the running effect assisted with the top Yukawa coupling enhanced by $\sin^{-1}\beta$.
%However, such a region is excluded by the tachyonic mode and false vacuum.
%Overall, $m_{h_1}$ is reduced with $\tan\beta$ due to the suppressed MSSM contribution, however, 
%in the region around $(\lambda, \kappa)=(0.65, 0.15)$ it reaches 125 GeV
%because of the additional Higgs quartic coupling, 
%$\lambda^2\sin^2 2\beta$.
%The ratio $g^2_{ZZh_1}/g^2_{\rm SM}$ is reduced to $\sim 0.95$ 
%where $m_{h_1}$ is enhanced, 
%while in the other region almost the same as in the SM.
%It is interesting to observe that the region with the ameliorated fine-tuning 
%and enhanced $m_{h_1}$ is exactly the region favored by
%the expected size of $|m_S|^2$ in the mirage mediation.
%Table 3. shows examples of the mass spectrum with the same parameters as in Table 1 except for $\tan\beta=3$.
%All the Higgs masses increase in contrast to $\tan\beta = 5$ case
%except for $m_{h_1}$ at $(\kappa, \lambda)=(0.1, 0.3)$.
%$\mu$ becomes significantly heavy, however, changes in $m_{H_u}^2$ are limited.

\begin{table}[htbp]
\begin{center}
\begin{tabular}{|c|c|c|c|c|}
\hline
$(\lambda,\kappa)$ & $(0.10,0.40)$ &  $(0.40,0.10)$ & $(0.70,0.11)$\\ \hline \hline
$m_{h1}$ & 113 GeV   & 79 GeV        & 65 GeV \\ \hline
$m_{h2}$ & 1136 GeV  & 126 GeV         & 130 GeV \\ \hline
$m_{h3}$ & 1209 GeV  & 1222 GeV        & 1228 GeV \\ \hline
$m_{a1}$ & 420 GeV   & 138 GeV        & 121 GeV \\ \hline
$m_{a2}$ & 1205 GeV  & 1221 GeV        & 1227 GeV \\ \hline
$g_{ZZh_1}^2/g^2_{\rm SM}$ & 1.00 & 0.24  & 0.13 \\ \hline

$m_{Hd}^2$ & $1.39 \times 10^6 \mathrm{GeV}^2$ 
        &  $1.39 \times 10^6 \mathrm{GeV}^2$
        & $1.39 \times 10^6 \mathrm{GeV}^2$ \\ \hline
$m_{Hu}^2$ & $4.91 \times 10^4 \mathrm{GeV}^2$  
        & $2.00 \times 10^4 \mathrm{GeV}^2$
        & $1.97 \times 10^4 \mathrm{GeV}^2$ \\ \hline
$m_{S}^2$ & $-6.23 \times 10^5 \mathrm{GeV}^2$ 
       &  $8.11 \times 10^2 \mathrm{GeV}^2$
       & $4.85 \times 10^3 \mathrm{GeV}^2$ \\ \hline
$m_{\tilde{t}_{1}}$ &  822 GeV &  822 GeV &  822 GeV\\ \hline
$m_{\tilde{t}_{2}}$ & 847 GeV & 847 GeV & 847 GeV \\ \hline
%$m_{\tilde{b}_{1}}$ & 1137.9 GeV \\ \hline
%$m_{\tilde{b}_{2}}$ & 1277.3 GeV \\ \hline
$\mu$ & 146 GeV & 228 GeV & 232 GeV \\ \hline
\end{tabular}
\caption{Spectra for $(\lambda, \kappa)=(0.10,0.40)$, $(0.40,0.10)$ and $(0.70,0.11)$
with $\tan\beta=5$ and $M_0 = 1200$GeV.}
\end{center}
\end{table}

%For completeness, we increase $M_0$ to $M_0 = 1500$ GeV and $M_0 = 1700$ GeV with the same $\tan\beta$ in Fig. 4 and 5, respectively. The mass spectrum shifts upwards with the same qualitative behavior as in Fig 3. The region of optimum fine-tuning becomes narrower than in the case with $M_0 = 1200$ GeV, however, $\mu$ can be as heavy as $800$ GeV. The corresponding mass spectra are listed in Table 4. and 5, respectively.

When we increase $\tan \beta$, $|\mu|$ decreases 
as expected from the rough estimation, 
$|\mu| \sim M_0/\tan \beta$ in Eq.~(\ref{eq:app-1}).
For example, the value  $\tan \beta = 10$ 
leads $\mu = 100$ GeV or less for $M_0\simeq 1$ TeV,
which is excluded by the chargino mass bound by LEPII.
The value of $|m_S|$ also decreases as $\tan \beta$ 
increases.
In opposite view, this means that tuning of $|m_S|$ (or $\delta c^{({\rm loop})}_i$)
increases to obtain large $\tan\beta$ for fixed $(\lambda, \kappa)$
%KO>
 and small $\tan\beta$ is favored in our scenario.
%<KO
%The lightest CP-even Higgs mass $m_{h_1}$ increases 
%with $\tan\beta$ due to the MSSM tree-level contribution,
%however saturates around $\tan\beta \sim 10$ and 
%$m_{h_1}\approx 125$ GeV can not be reached with $M_0=1-3$ TeV.
%If we increase $M_0$ beyond this range, the fine-tuning deteriorates
%%KO>
%into ${\it O}(1)$ \% or more.
%%<KO
%%%On the other hand, $|\mu|$ and $|m_s|$ become larger 
%%%for smaller $\tan \beta$.

Finally we comment on the fermionic sector.
The singlino mass is estimated as 
$2\kappa \mu / \lambda$.
Since we have $\mu$ around $200 - 500$ GeV, 
the singlino can also be light.
Note that the region with small $\kappa/\lambda$ is 
excluded by the appearance of the tachyonic modes and/or the false vacua.
Thus, we can not lead the singlino much lighter than 
the higgsino.
Also, large $\kappa/\lambda$ leads large $|m_S|$, 
which could not be derived in our TeV scale mirage scenario.
Thus, singlino much heavier than the higgsino 
is disfavored.
Thus, both masses of the higgsino and singlino 
would be of the same order.
Since all of gaugino masses as well as squark and slepton masses 
are much heavier, the lightest superparticle would be 
a linear combination between the higgsino and singlino, 
depending on $\kappa/\lambda$.
The gravitino is quite heavy such as $m_{3/2} \sim 4\pi^2 M_0$, 
as already known in the mirage mediation mechanism 
\cite{Choi:2004sx,Choi:2005uz,Endo:2005uy}.

%%KO
The lightest neutralino in this model behaves like the linear combination between the 
higgsino and bino in the MSSM and could reproduce the observed abundance of
the cold dark matter assuming the thermal relic saturates it
\cite{Belanger:2005kh}. 
While the extra Higgs bosons play an important
role in the direct detection of them,
which could be significantly different from the results in the MSSM \cite{Cerdeno:2004xw}. 
The detailed study of the phenomenology
of the TeV scale mirage mediation in the NMSSM is interesting for its distinct mass spectrum, however,
beyond the scope of this work \cite{phenomenology}.

%\section{Phenomenology}

\section{Conclusion}

We have studied the NMSSM with the TeV scale mirage mediation.
The region with large $\kappa/\lambda$ requires 
a large value of $|m_S|$ to satisfy the 
stationary conditions of the Higgs potential.
Such a large value could not be realized in 
our TeV scale mirage scenario 
and therefore such parameter region is disfavored.
In the favored region, it is found that we can realize 
%$\mu (=\lambda \langle S \rangle ) \sim 200$ GeV, 
$|m_{H_u}| \sim 200$ GeV, 
while other masses are heavy such as $1$ TeV.
Then, the fine-tuning problem is ameliorated.
The cancellation between the effective $\mu$-term and the down-type Higgs soft
mass reduces the sensitivity of $\mu$ to $m_Z$ and $\mu={\cal O}(500)$ GeV is possible without significantly deteriorating the fine-tuning.
%KO>
The mixing between the light doublet and the singlet is suppressed by $(\kappa/\lambda)\tan^{-1}\beta$.
%<KO
%KO>
%The lightest CP-even Higgs mass reaches $125$ GeV at small $\tan\beta$
%where the mixing between the doublets and the singlet vanishes.
For small $\tan\beta$ the lightest CP-even Higgs is mainly the doublet and
 its mass reaches $125$ GeV without suppression by the mixing.
%<KO
%The lightest CP-even Higgs mass is obtained as 
%$m_{h_1} = 115 - 130$ GeV for $M_0 = 1200-1500$ GeV.
%
When we increase $M_0$, the Higgs mass $m_{h_1}$ slowly increases, 
and also the value of $|\mu|$ increases.
However, the required fine-tuning is still mild e.g. 
for $M_0 = 1700$ GeV .

The coupling between the lightest CP-even Higgs boson 
and the $Z$ boson is almost the same as one in the standard model
for small $\tan\beta$, however, can decrease to ${\it O}(10)$ \% 
for moderate $\tan\beta$
%KO>
 where the lightest CP-even Higgs is mainly singlet.
If the lightest CP-even Higgs escapes the LEPII bound,
the second lightest CP-even Higgs boson 
could be the boson observed in LHC.
The mass of the second lightest CP-even Higgs boson 
depends on $\kappa$ and $\lambda$, 
and it can be light in the parameter region favored in our scenario.
%<KO
The heaviest CP-even Higgs mass as well as 
the heaviest CP-odd and charged ones is of ${\cal O}(M_0)$.
The lightest CP-odd Higgs can also be light.
Thus, the Higgs sector has a rich structure.

In our scenario, the higgsino is light compared with three 
gauginos.
In addition, the singlino is also light.
Both masses of the higgsino and the singlino are of the same order.
Then, the lightest superparticle is a linear combination 
between the higgsino and singlino.
Such a neutralino sector and Higgs sector would lead to 
several phenomenologically interesting aspects.

\vskip 1cm
{\bf Note added}

While this work was being completed, we received
Ref.~\cite{Asano:2012sv}, which also considered 
relevant aspects.

\section*{Acknowledgments}

T.~K. is supported in part by a Grant-in-Aid for Scientific Research No.~20540266 and the Grant-in-Aid for the 
Global COE Program ``The Next Generation of Physics, Spun from Universality and Emergence'' from the Ministry of 
Education, Culture, Sports, Science and Technology of Japan. 
K.~O. is supported in part by a Grant-in-Aid for Scientific Research No.~21740155 and No.~18071001 from the MEXT of Japan. 
T.~S. is a Yukawa Fellow and his work is partially supported by the Yukawa Memorial Foundation, 
a Graint-in-Aid for Young Scientists (B) No.~23740190 and the Sasakawa Scientific Research Grant from the Japan Science Society.
%%以下はYITP support のため。
Numerical computation in this work was partly carried out at the Yukawa Institute Computer Facility.

\appendix

\section{Soft SUSY breaking terms}
\label{app:soft}

Here we give explicitly soft SUSY breaking terms induced 
by the mirage mediation mechanism in the NMSSM.

In the mirage mediation, 
the soft parameters at the scale  just  below  $M_{GUT}$ are given by
\begin{eqnarray}
M_a(M_{GUT}) &=& M_0 + \frac{m_{3/2}}{8\pi^2} b_a g_a^2, \nonumber \\
A_{ijk}(M_{GUT}) &=& (c_i + c_j + c_k)M_0 - (\gamma_i + \gamma_j + \gamma_k)\frac{m_{3/2}}{8\pi^2}, \nonumber \\
m_i^2(M_{GUT}) &=& c_i M_0^2 - \dot{\gamma_i}(\frac{m_{3/2}}{8\pi^2})^2
				- \frac{m_{3/2}}{8\pi^2} M_0 \theta_i,
\end{eqnarray}
where
\begin{eqnarray}
b_a &=& -3\mathrm{tr}(T_a^2(\mathrm{Adj})) + \sum_i \mathrm{tr}(T_a^2(\phi^i)), \nonumber \\
\gamma_i &=& 2\sum_a g_a^2 C_2^a(\phi^i) - \frac{1}{2} \sum_{jk} |y_{ijk}|^2, \nonumber \\
\theta_i &=& 4\sum_a g_a^2 C_2^a(\phi^i) - \sum_{jk} a_{ijk} |y_{ijk}|^2, \nonumber \\
\dot{\gamma_i} &=& 8\pi^2 \frac{d\gamma_i}{d \ln \mu_R}.
\end{eqnarray}
Here, $T_a^2(\mathrm{Adj})$ and $T_a^2(\phi^i)$  denote 
Dynkin indices of the adjoint representation and the representation 
of matter fields $\phi^i$.
We have assumed
$\omega_{ij}=\Sigma_{kl}y_{ijk} y_{jkl}^*$ to be diagonal.

Within the framework of the NMSSM, the $\beta$-function coefficients, anomalous 
dimensions and other coefficients in the above equations are obtained as 
\begin{eqnarray}
b_3 &=& -3,\  b_2 = 1,\  b_1 = 11, \nonumber \\
\gamma_{H_u} &=& \frac{3}{2}g_2^2 + \frac{1}{2}g_1^2 - 3y_t^2 - \lambda^2,  \nonumber \\
\gamma_{H_d} &=& \frac{3}{2}g_2^2 + \frac{1}{2}g_1^2  - \lambda^2,  \nonumber \\
\gamma_{S} &=& -2\kappa^2 - 2\lambda^2,  \nonumber \\
\gamma_{Q_a} &=& \frac{8}{3}g_3^2 + \frac{3}{2}g_2^2 + \frac{1}{18}g_1^2 - (y_t^2 + y_b^2) \delta_{3a},  \nonumber \\
\gamma_{U_a} &=& \frac{8}{3}g_3^2 + \frac{8}{9}g_1^2 - 2y_t^2 \delta_{3a},  \nonumber \\
\gamma_{D_a} &=& \frac{8}{3}g_3^2 + \frac{2}{9}g_1^2 - 2y_b^2 \delta_{3a},  \nonumber \\
\gamma_{L_a} &=& \frac{3}{2}g_2^2 + \frac{1}{2}g_1^2 - y_{\tau}^2 \delta_{3a}, \nonumber \\
\gamma_{E_a} &=& 2g_1^2 - 2y_{\tau}^2 \delta_{3a},
\end{eqnarray}

\begin{eqnarray}
\theta_{H_u} &=& 3g_2^2 + g_1^2 - 6y_t^2 a_{H_uQ_3U_3^c} -2\lambda^2a_{H_uH_dS},  \nonumber \\
\theta_{H_d} &=& 3g_2^2 + g_1^2 - 6y_b^2 a_{H_dQ_3D_3^c} - 2y_{\tau}^2 a_{H_dL_3E_3^c} -2\lambda^2a_{H_uH_dS},  \nonumber \\
\theta_{S} &=& -2\lambda^2a_{H_uH_dS} -\kappa^2 a_{SSS},  \nonumber \\
\theta_{Q_a} &=& \frac{16}{3}g_3^2 + 3g_2^2 + \frac{1}{9}g_1^2 - 2(y_t^2 a_{H_uQ_3U_3^c} + y_b^2 a_{H_dQ_3D_3^c}) \delta_{3a}, \nonumber \\
\theta_{U_a} &=& \frac{16}{3}g_3^2 + \frac{16}{9}g_1^2 - 4y_t^2 a_{H_uQ_3U_3^c}\delta_{3a},  \nonumber \\
\theta_{D_a} &=& \frac{16}{3}g_3^2 + \frac{4}{9}g_1^2 - 4y_b^2 a_{H_dQ_3D_3^c}\delta_{3a}, \nonumber \\
\theta_{L_a} &=& 3g_2^2 + g_1^2 - 2y_{\tau}^2 a_{H_dL_3E_3^c}\delta_{3a}, \nonumber \\
\theta_{E_a} &=& 4g_1^2 - 4y_{\tau}^2 a_{H_dL_3E_3^c}\delta_{3a},
\end{eqnarray}

\begin{eqnarray}
\dot{\gamma}_{H_u} &=& \frac{3}{2}g_2^4 + \frac{11}{2}g_1^4 - 3y_t^2 b_{y_t} -\lambda^2 b_\lambda,  \nonumber \\
\dot{\gamma}_{H_d} &=& \frac{3}{2}g_2^4 + \frac{11}{2}g_1^4 - 3y_b^2 b_{y_b} - y_{\tau}^2 b_{y_\tau} -\lambda^2 b_\lambda, \nonumber \\
\dot{\gamma}_{S} &=& -2\kappa^2 b_\kappa - 2\lambda^2 b_\lambda,  \nonumber \\
\dot{\gamma}_{Q_a} &=& -8g_3^4 + \frac{3}{2}g_2^4 + \frac{11}{18}g_1^4 - (y_t^2 b_{y_t} + y_b^2 b_{y_b}) \delta_{3a},  \nonumber \\
\dot{\gamma}_{U_a} &=& -8g_3^4 + \frac{88}{9}g_1^4 - 2y_t^2 b_{y_t} \delta_{3a},  \nonumber \\
\dot{\gamma}_{D_a} &=& -8g_3^4 + \frac{22}{9}g_1^4 - 2y_b^2 b_{y_b} \delta_{3a},  \nonumber \\
\dot{\gamma}_{L_a} &=& \frac{3}{2}g_2^4 + \frac{11}{2}g_1^4 - y_{\tau}^2 b_{y_\tau} \delta_{3a}, \nonumber \\
\dot{\gamma}_{E_a} &=& 22g_1^4 - 2y_{\tau}^2 b_{y_\tau} \delta_{3a},
\end{eqnarray}
where
\begin{eqnarray}
b_{y_t} &=& -\frac{16}{3}g_3^2 -3g_2^2 - \frac{13}{9}g_1^2 + 6y_t^2 + y_b^2 + \lambda^2,  \nonumber \\
b_{y_b} &=& -\frac{16}{3}g_3^2 -3g_2^2 - \frac{7}{9}g_1^2 + y_t^2 + 6y_b^2 + y_\tau^2 + \lambda^2, \nonumber \\
b_{y_\tau} &=& -3g_2^2 - 3g_1^2 + 3y_b^2 + 4y_\tau^2 + \lambda^2,  \nonumber \\
b_{\kappa} &=&  6\lambda^2 + 6\kappa^2,  \nonumber \\
b_{\lambda} &=& -3g_2^2 - g_1^2 + 3y_t^2 + 3y_b^2 + y_\tau^2 + 4\lambda^2 + 2\kappa^2.
\end{eqnarray}
Here, $Q_a, U_a, D_a, L_a,$  and  $E_a$ denote left-handed quark, right-handed up-sector quark, 
right-handed down-sector quark, left-handed lepton, and right-handed lepton fields, respectively, 
and the index $a$ denotes the generation index.
We have included effects due to Yukawa couplings, 
$y_t, y_b,$ and $y_\tau$, only for the third generations.

%%%%%%%%%%%%%%%%%%%%%%%
% Shimomura added the following figures
%%%%%%%%%%%%%%%%%%%%%%%

\begin{figure}[h]
\begin{center}
\begin{tabular}{l @{\hspace{10mm}} r }
\includegraphics[height=55mm]{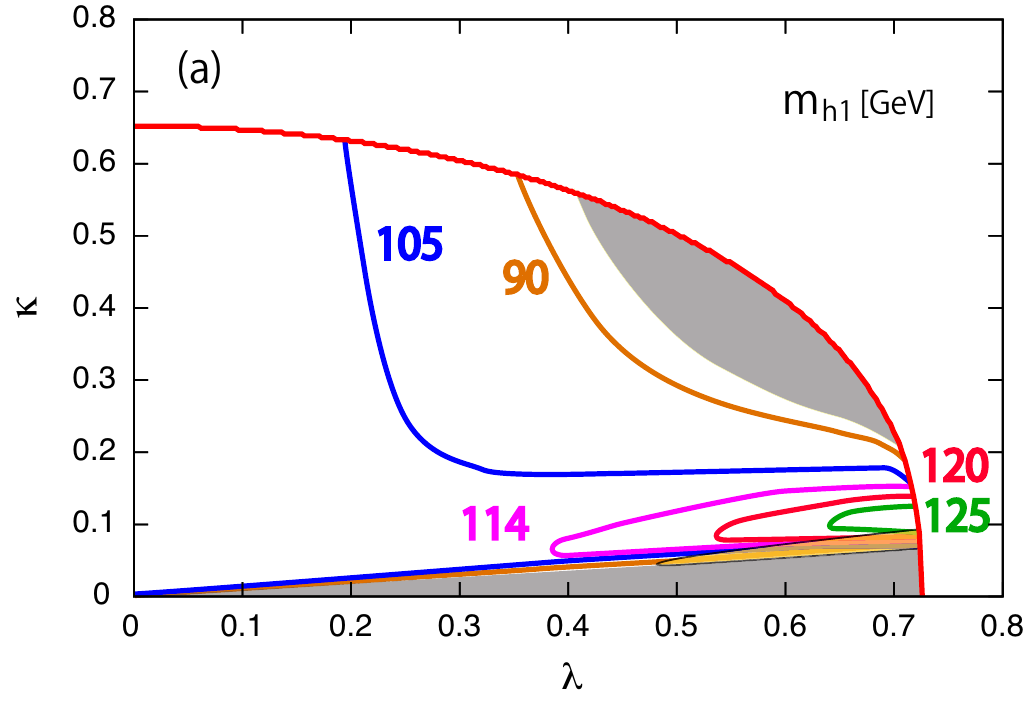} &
\includegraphics[height=55mm]{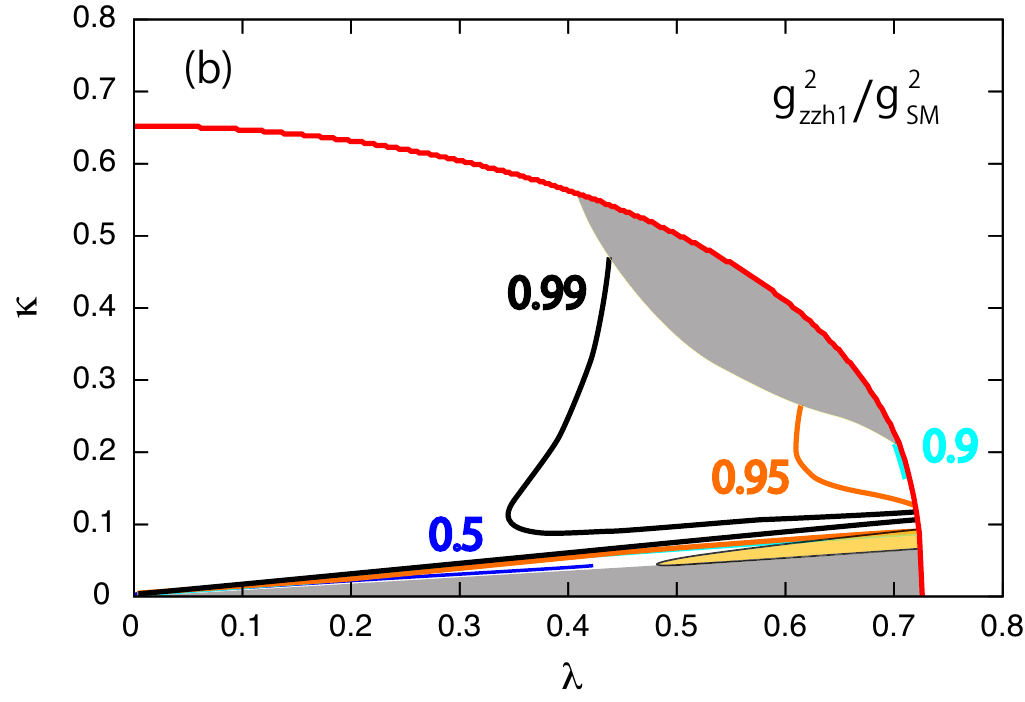} \\
\rule{0cm}{10mm} & \\
\includegraphics[height=55mm]{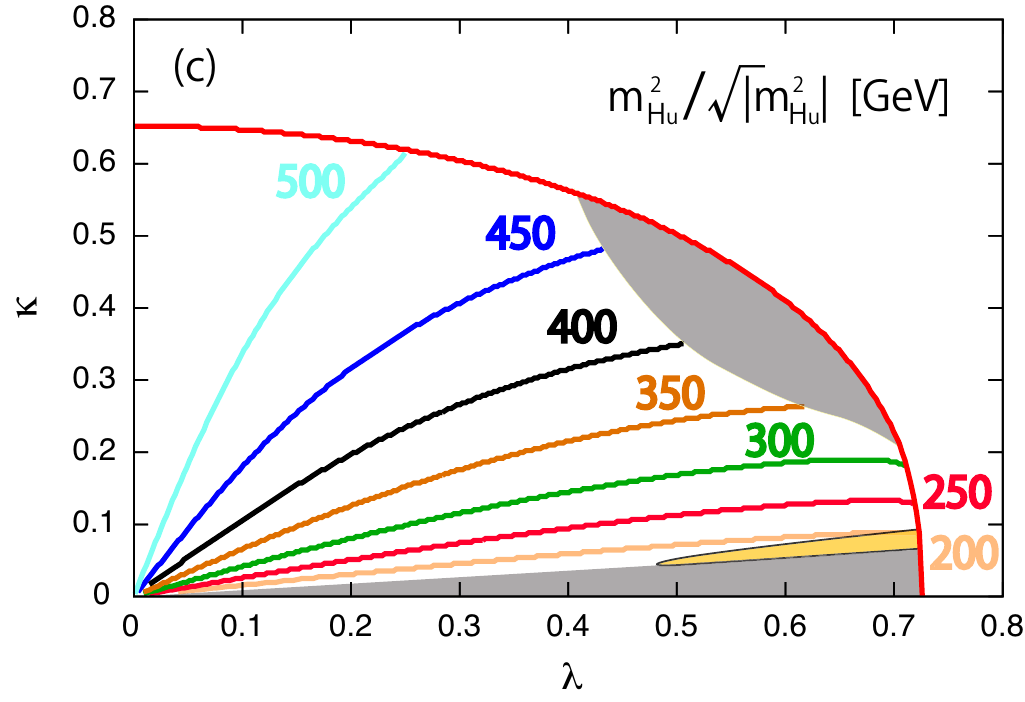} &
\includegraphics[height=55mm]{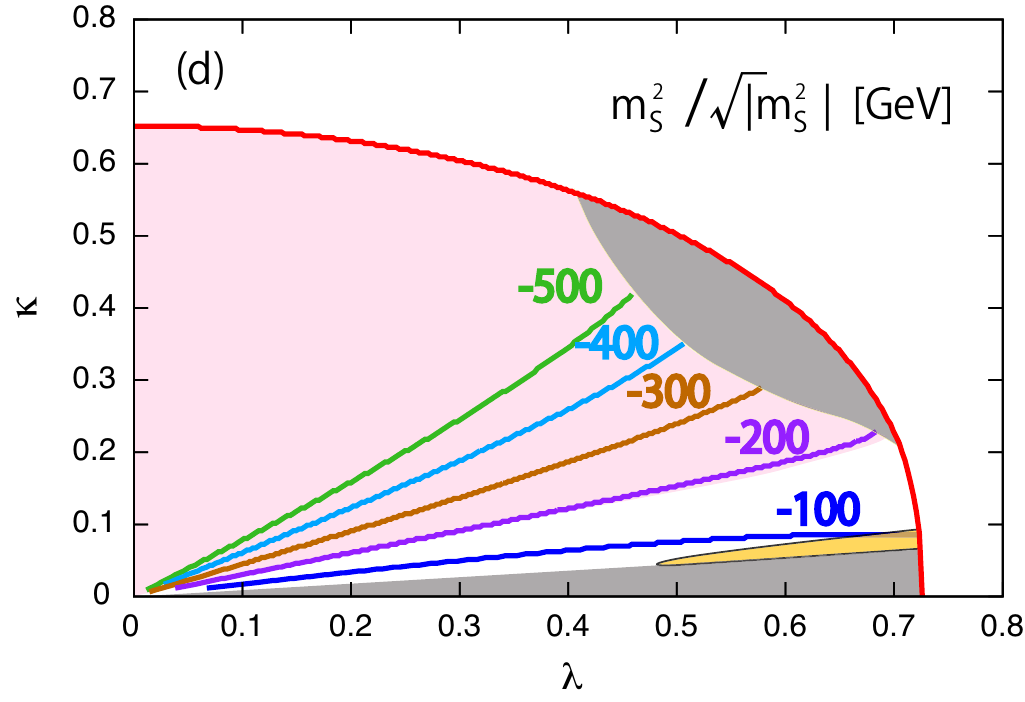} \\
\rule{0cm}{10mm} & \\
\includegraphics[height=55mm]{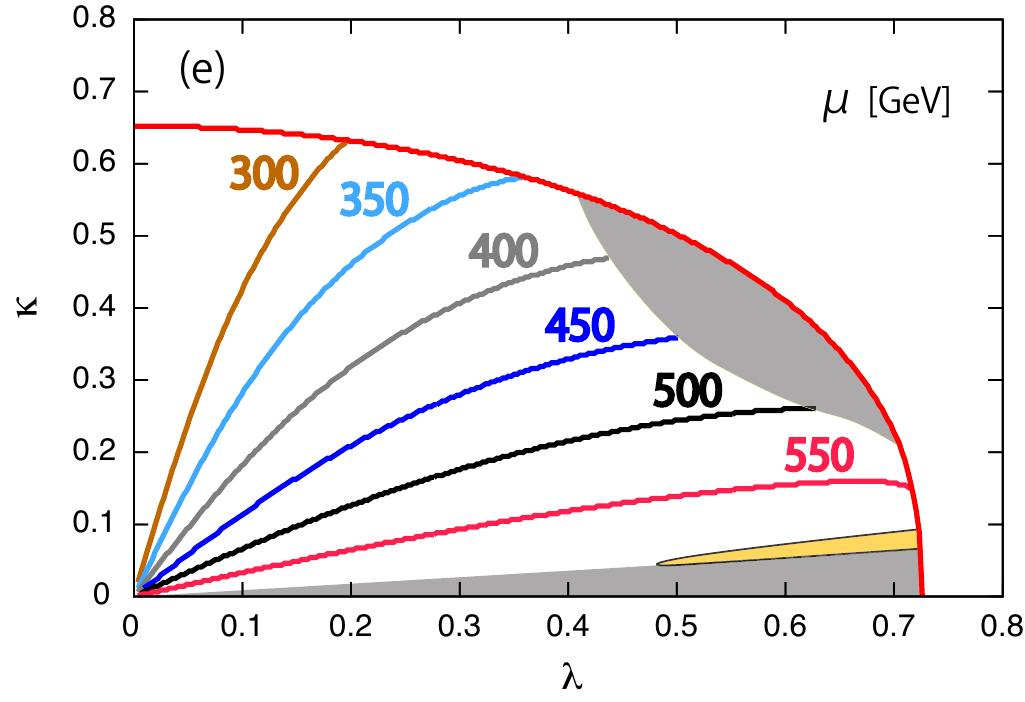} &
\includegraphics[height=55mm]{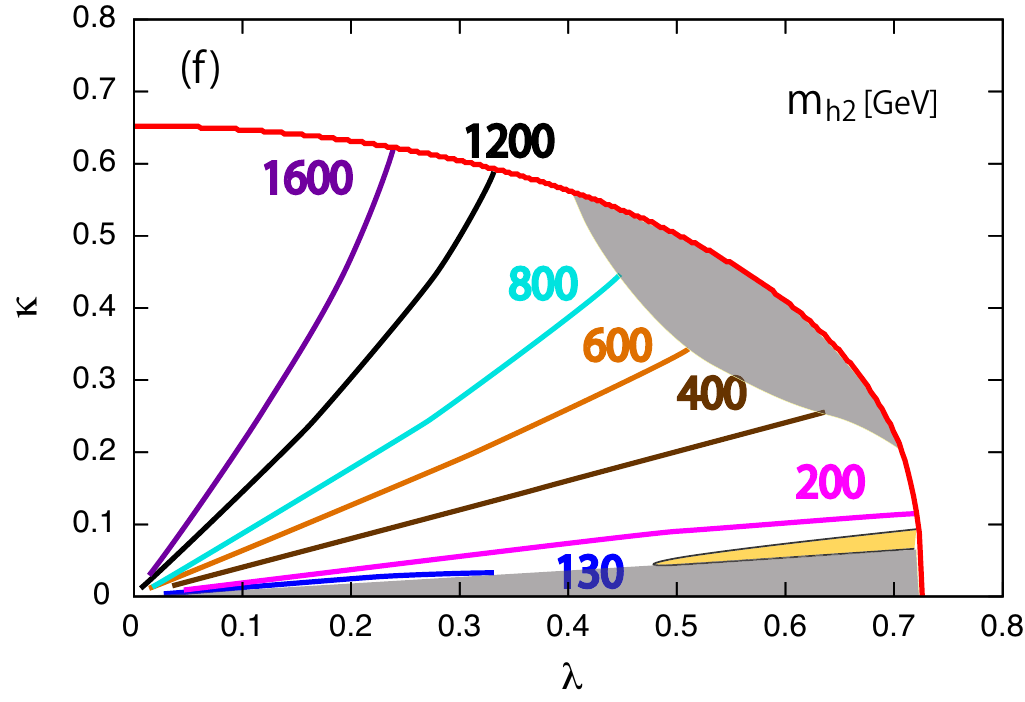}
\end{tabular}
\caption{Constraints and SUSY breaking parameters.
We use $M_0 = 1700GeV, \alpha=2, c_{Hd}=1, c_{Hu}=0, c_S=0, c_{\tilde{Q}}=0.5, c_{\tilde{t}}=0.5$
and $\tan\beta=3 , A_\kappa=-100\mathrm{GeV}$.}
\end{center}
\end{figure}

\begin{figure}[h]
\begin{center}
\begin{tabular}{l @{\hspace{10mm}} r }
\includegraphics[height=55mm]{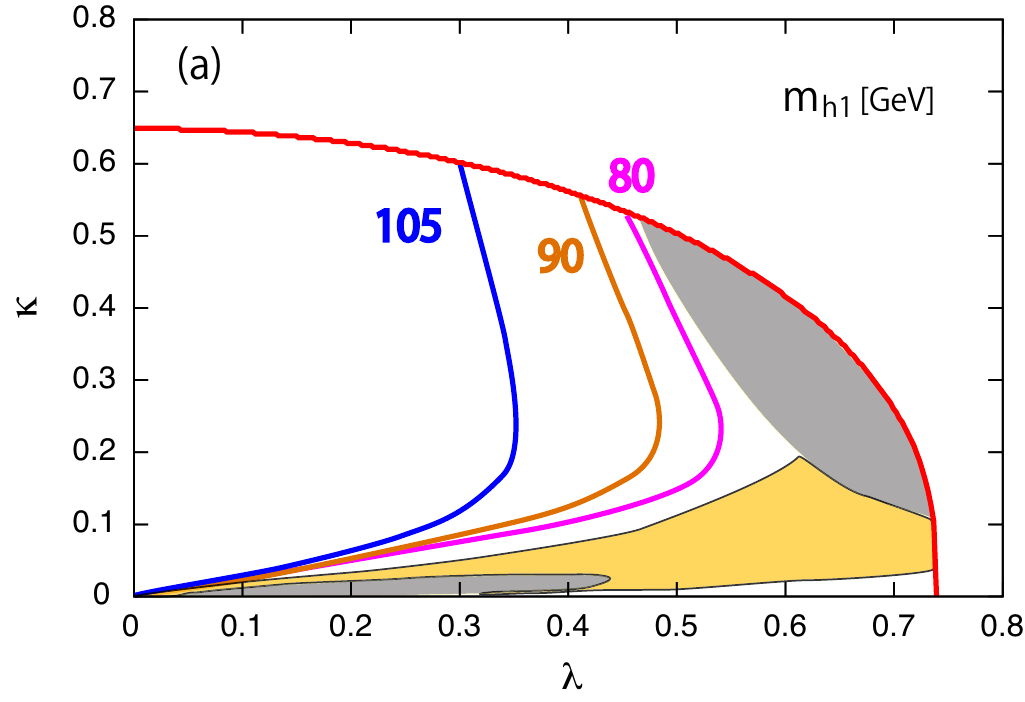} &
\includegraphics[height=55mm]{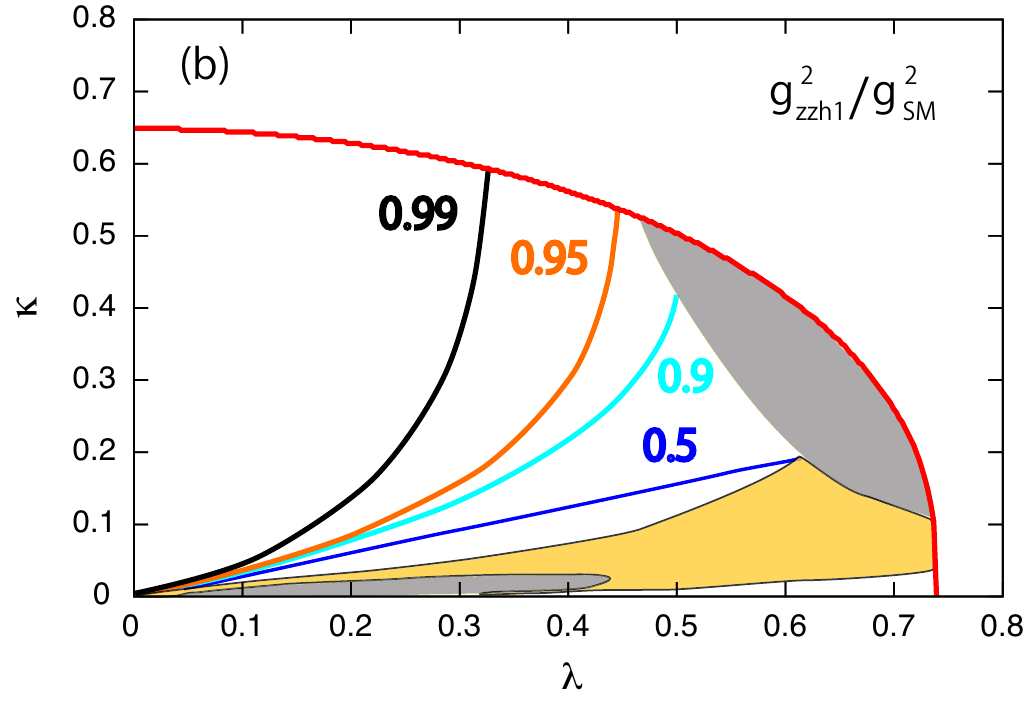} \\
\rule{0cm}{10mm} & \\
\includegraphics[height=55mm]{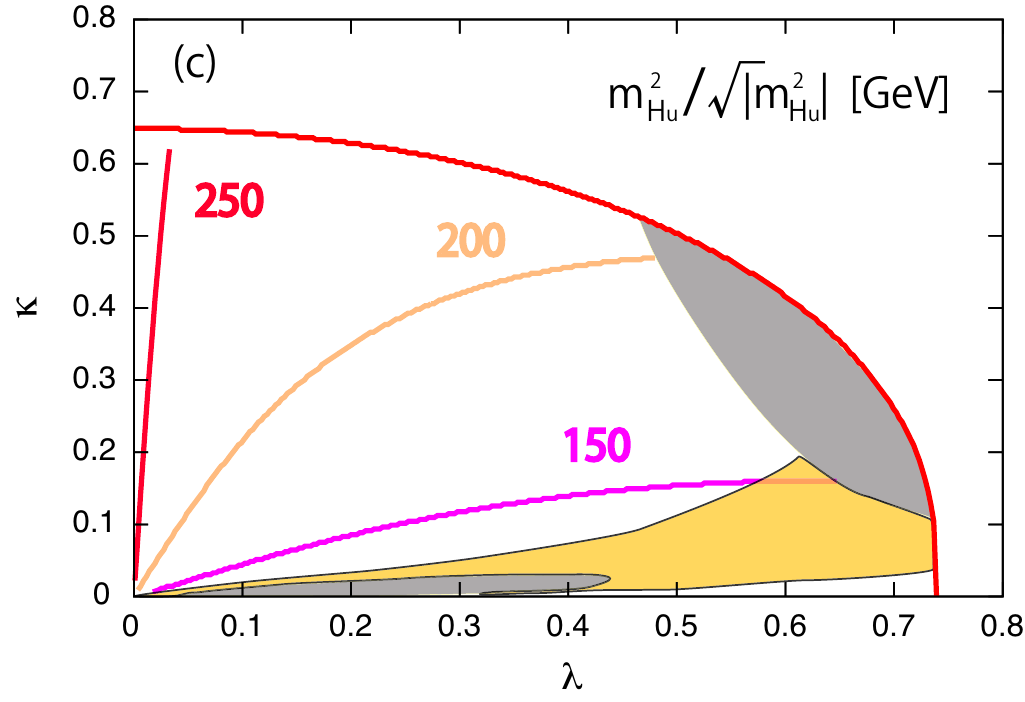} &
\includegraphics[height=55mm]{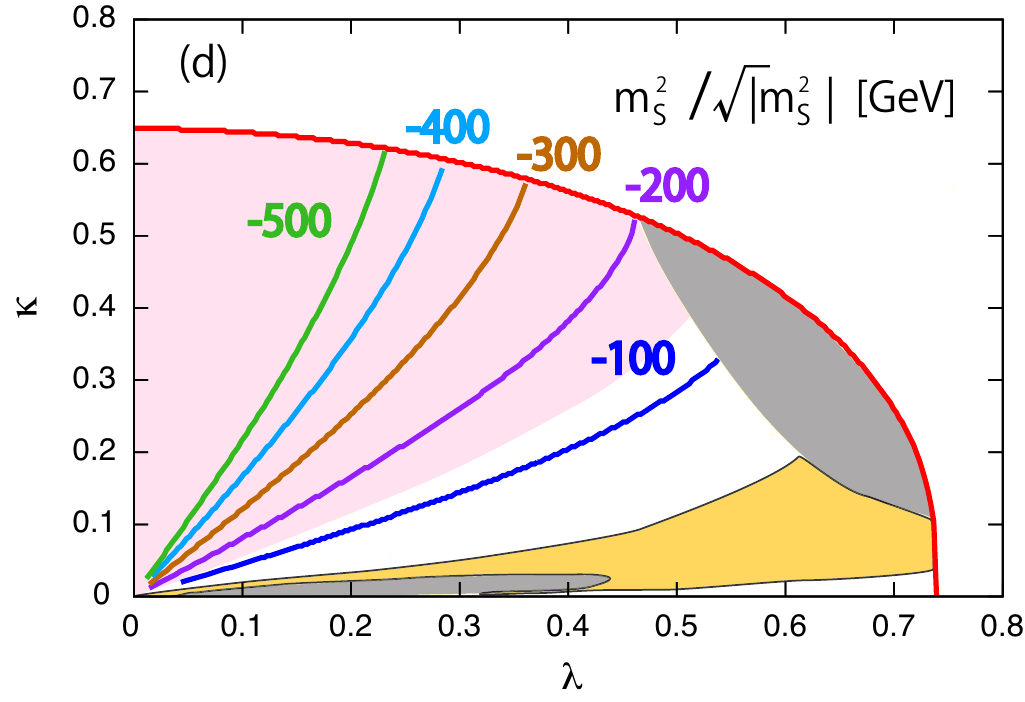} \\
\rule{0cm}{10mm} & \\
\includegraphics[height=55mm]{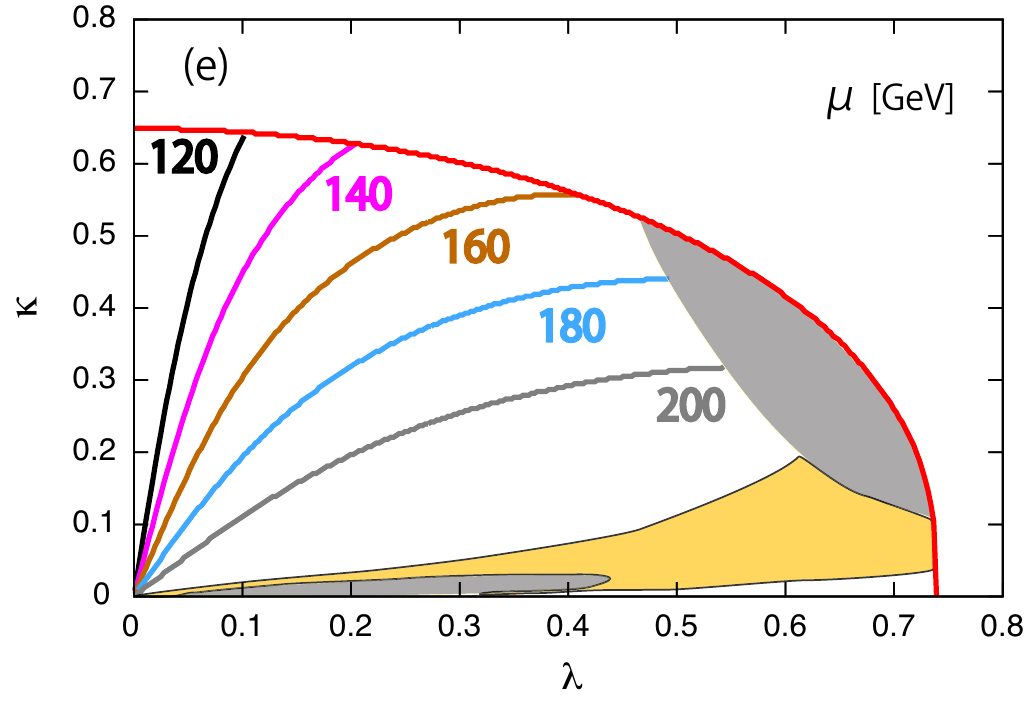} &
\includegraphics[height=55mm]{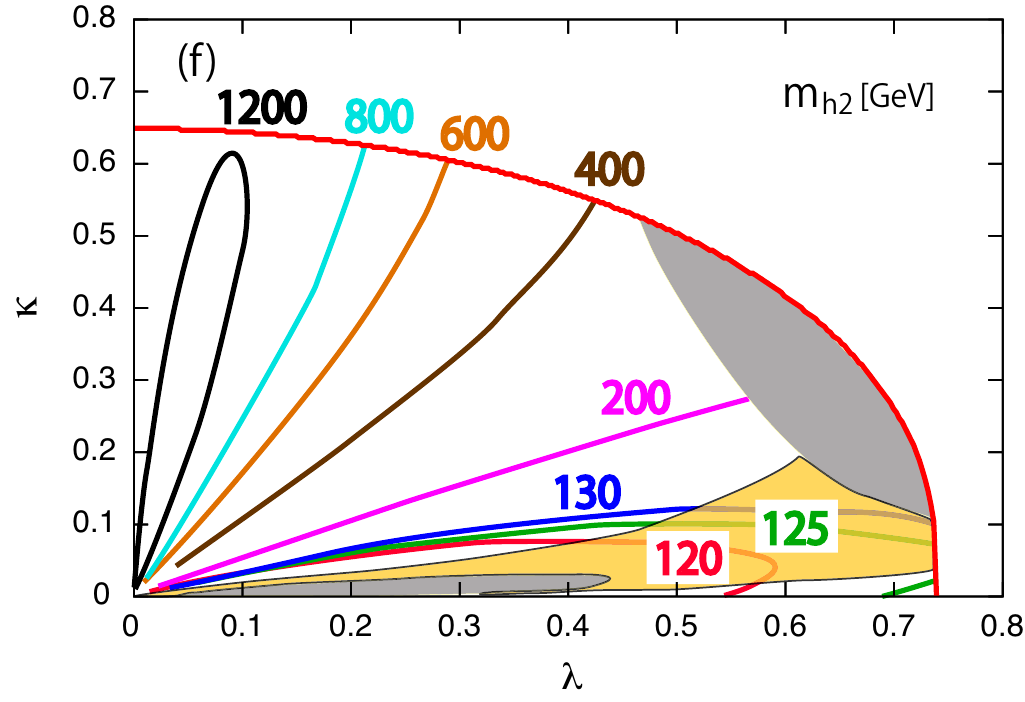}
\end{tabular}
\caption{Constraints and SUSY breaking parameters.
We use $M_0 = 1200GeV, \alpha=2, c_{Hd}=1, c_{Hu}=0, c_S=0, c_{\tilde{Q}}=0.5, c_{\tilde{t}}=0.5$
and $\tan\beta=5 , A_\kappa=-100\mathrm{GeV}$.}
\end{center}
\end{figure}

\clearpage

%%%%%%%%%%
% end Shimomura
%%%%%%%%%%

\end{document}